\documentclass[12pt]{iopart}
\usepackage{graphicx}
\usepackage{color}
\usepackage{setstack}
\usepackage{amstext}
\usepackage{amssymb}
\usepackage{fixltx2e}

\definecolor{orange}{rgb}{1,0.5,0}

\begin{document}
\title[2D capillary wetting]{Wetting in a two-dimensional capped capillary. Part I: Wetting temperature and capillary prewetting.}
\author{P Yatsyshin$^1$ and N Savva$^{2}$ and S Kalliadasis$^1$}
\address{$^1$ Department of Chemical Engineering, Imperial College London,
London SW7 2AZ, United Kingdom}
\address{$^2$ School of Mathematics, Cardiff University, Cardiff, CF24 4AG, United Kingdom}
\ead{\mailto{p.yatsyshin@imperial.ac.uk},\mailto{SavvaN@cardiff.ac.uk},
\mailto{s.kalliadasis@imperial.ac.uk}}
\begin{abstract}
In this two-part study we investigate the phase behaviour of a fluid
spatially confined in a semi-infinite rectangular pore formed by three
orthogonal walls and connected to a reservoir maintaining constant values of
pressure and temperature in the fluid. Far from the capping wall this
prototypical two-dimensional system reduces to a one-dimensional slit pore.
However, the broken translational symmetry leads to a wetting behavior
strikingly different from that of a slit pore. Using a realistic model of an
atomic fluid with long-ranged Lennard-Jones fluid-fluid and fluid-substrate
interactions, we present for the first time detailed computations of full
phase diagrams of two-dimensional capped capillaries. Our analysis is based
on the statistical mechanics of fluids, in particular density functional
theory. We show the existence of capillary wetting temperature, which is a
property of the pore, and relatively to the fluid temperature determines
whether capillary condensation is a first-order or a continuous phase
transition. We also report for the first time a first-order capillary wetting
transition, which can be preceded by a first-order capillary prewetting. A
full parametric study is undertaken and we support our findings with
exhaustive examples of density profiles, adsorption and free energy
isotherms, as well as full phase diagrams.
\end{abstract}
\submitto{\JPCM}
\maketitle
\section{Introduction}

The origins of the field of wetting are often associated with the seminal
works of Cahn \cite{Cahn77} and Ebner and Saam \cite{EbnerSaam}, where a new
class of phase transitions caused by the non-uniformity of fluids in the
microscale had been envisaged and studied for the first time. Since then, the
field has seen a rapid growth, especially over the last two decades or so,
with a multitude of experimental and theoretical studies which have clearly
defined wetting as an outstanding cross-disciplinary field driving and
building upon progress in statistical physics, simulations and even
hydrodynamics and quantum mechanics \cite{BonnEtAlRevModPhys09, Squires05,
Bonn01}.

The usual macroscopic description of fluid interfaces based on surface
tensions becomes rather limited at the nano-scale where the fluid is
non-uniform \cite{DutkaNapirowkyDietrichJChemPhys12}. In fact a satisfactory
model for the fluid and associated phenomena such as wetting and surface
phase transitions at the nano-scale should account for the molecular
interactions in the system. Popular theoretical approaches for such phenomena
include phenomenological Landau or Van der Waals theories, interfacial
Hamiltonians, rigorous statistical mechanical approaches, such as mean field
lattice models and density functional (DF) theories, and simulations
\cite{Evans,Plischke,Hansen,WuAIChE06}.

DF theory in particular allows one to obtain detailed information about the
structure of the fluid, but all DF models are mainly restricted to numerical
studies and also cannot account for the fluctuation effects. Models based on
effective Hamiltonians, on the other hand, are often amenable to analytic
investigations, and may include the effects of thermal fluctuations in the
system. When coupled with a renormalization technique, such models can offer
a powerful tool for obtaining rigorous results, such as critical exponents
and universality classes, see, e.g. references
\cite{Parry07,TasDietPRL06,ParryPRL00,ParryRasconWoodPRL99,CardyNightingalePhysRevB83}.
However, unlike DF models, effective Hamiltonians cannot provide details of
the microscopic fluid structure as they \emph{a priori} assume the existence
of interfaces, often a particular shape. The limitations of this approach
have been discussed in reference~\cite{Dietrich}.

In the present study we employ a microscopic mean field DF approach. The
details of the formalism, limits of applicability, as well as references to
rigorous proofs can be found in the reviews of Wu~\cite{WuAIChE06} and
Evans~\cite{EvansSchool09}. Starting from reasonable assumptions about the
nature of fluid-fluid and fluid-substrate forces one can obtain theoretically
consistently all the characteristics of an adsorption process, such as
surface tensions, contact angles, interfaces, etc. Sophisticated DF models
offer a computationally inexpensive alternative to molecular dynamic
simulations, and often allow to reproduce experimental results. Recent
developments include the studies of adsorption of Argon, Neon and Xenon on
planar substrates of various compositions, see references~\cite{Zeng10, Yu10,
Sart09, AncilottoToigoJChemPhys00}. DF theories also provide a convenient
bridge between the nano- (molecular simulation) and micro- (experimentally
accessible) scales. Since DF calculations can be implemented without
truncating the tails of interaction potentials \cite{Yatsyshin2012}, they
have a principal advantage over simulations and also provide a way to verify
analytically predicted critical exponents \cite{Yatsyshin2013,Roth11}. At the
same time, the relatively low cost of DF calculations allows to explore the
parameter space of models, obtaining complete phase diagrams and possibly
uncovering qualitatively new phenomena.

A DF model approximates the additive free energy of the fluid, which consists
of two main parts: the fluid-fluid and the fluid-substrate contributions. The
fluid-fluid free energy should account for long-ranged attractive and
short-ranged repulsive forces between fluid molecules. The fluid-substrate
interactions are typically modelled by considering the substrate as an inert
spectator phase and accounting for its effect on the fluid by an
appropriately chosen external potential, which enters the expression for the
free energy.

Finally, DF models can be extended to dynamic systems, using e.g. models A or
B of the general dynamic universality classes~\cite{Hohenberg77}, which
allows to capture the characteristics of the system as it relaxes slowly to
its equilibrium, see e.g. references
\cite{GoddardPRL12,GoddardJPhysCondMat13,ArcherEvans,MarTar00}.

In the present study we consider the phase behaviour of a fluid confined in a
prototypical two-dimensional (2D) pore. We show how the dimensionality of the
problem dramatically affects the phenomenology of wetting. Investigations of
wetting on nano-structured substrates are interesting from a fundamental
statistical physics perspective, as confined fluid surfaces may exhibit
various kinds of first-order and continuous phase transitions
\cite{BinderAnnuRevMaterRes08,HerminghausBrinkmanSeemanAnnuRevMatRes08}.
Exciting new phenomena are mainly determined by the interplay of various
length and energy scales in the system, e.g., ranges and strengths of
fluid-fluid and fluid-substrate potentials, characteristic dimensions of
confining geometries, and even particle sizes. These parameters may act as
thermodynamic fields and lead, according to the Gibbs phase rule, to a
multitude of phase transitions and metastable states. Capillary phenomena
associated with nano-confinement provide a vivid manifestation of attractive
molecular forces, putting to a test our microscopic picture of matter
\cite{ChandlerWeeksScience83}. Active applied interest in chemical
engineering includes microfluidics
\cite{RauschDietAnnuRevMaterRes08,CraigheadNature06}, design of nano-scale
chemical reactors \cite{Dum09,Calvo09} and biomimetic surfaces
\cite{GouLiuPlantSci07} or surfaces with variable wetting properties
\cite{BerardinoDietrich2012}.

Consider a slit pore formed by two parallel planar walls of infinite area,
which are separated by a distance $H$. When the system is immersed in a large
reservoir filled with vapour at pressure $P$ (the temperature is assumed
fixed), simple thermodynamic considerations of the energy necessary to form
two wall-fluid interfaces lead to the possibility of two-phase coexistence
within the pore: between vapour and capillary-liquid phases (see, e.g.,
\cite{SaamJLowTempPhys09}). The transition between these phases is in fact
the shifted bulk liquid -- vapour transition (saturation). It is a
first-order phenomenon, where the pore becomes filled with a denser phase
(capillary-liquid) discontinuously, if the pressure in the reservoir becomes
equal to $P_{\text{v}}$, defined by the Kelvin equation:
\begin{equation}
\label{Kelvin}
P_{\text{v}}=P_{\text{l}}-\frac{2\sigma_{\text{lv}}\cos\Theta}{H}+\dots,
\end{equation}
where $P_{\text{l}}$ is the pressure at bulk saturation, $\sigma_{\text{lv}}$
is the liquid -- vapour surface tension and $\Theta$ is the macroscopic
contact angle at bulk coexistence.  The above expression neglects various
microscopic effects, the most important being the competition between
fluid-fluid and fluid-wall forces, which can be accounted for by, e.g., an
effective interfacial potential \cite{Dietrich}, or, more consistently,
within a DF approach \cite{Evans90}. A detailed description of surface
thermodynamics associated with this prototypical case of confinement can be
found in, e.g., reference \cite{EvMar87}.

\begin{figure}
\centering
    \includegraphics{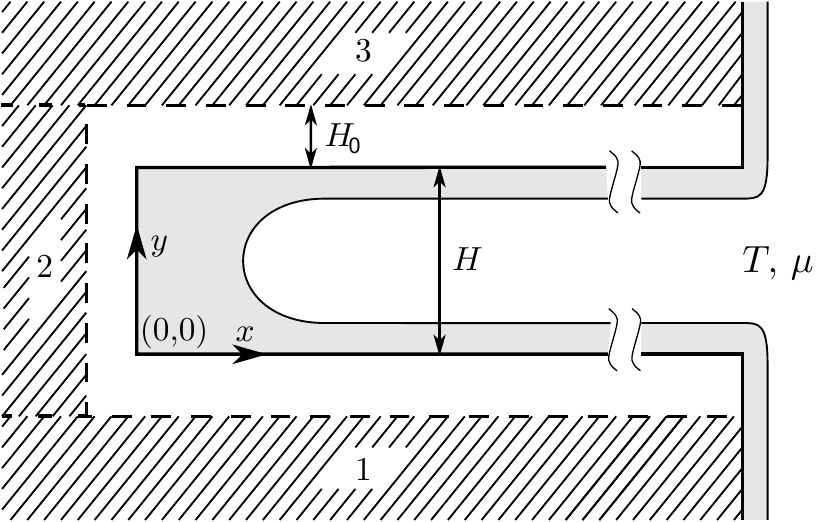}%
    \caption{Sketch of adsorption in the capped capillary of width $H$ opening into a reservoir of particles on the right
    at bulk temperature $T$ and chemical potential $\mu$.
    The gray area denotes an adsorbed capillary-liquid. The hatched areas denote
    parts of the substrate with parameters of pairwise LJ fluid-substrate
    interactions $\varepsilon_{\text{w}}$, $\sigma_{\text{w}}$. The substrate
    potential has a cutoff $H_0\geq\sigma_{\text{w}}$. The directions of the two axis
    $x$ and $y$ and the origin $\left(0,0\right)$ are also shown. Tildes denote
    the infinitely long separation of the capping wall on the left from the
    channel opening into the reservoir on the right.
     \label{FigOne}}%
\end{figure}

Now consider a \emph{capped capillary}, namely a slit pore closed at one end
by a third wall, as shown in figure \ref{FigOne}. On one hand, one expects
the transition to a state, where the pore becomes filled with
capillary-liquid from vapour -- capillary condensation (CC) -- to be of first
order, since the capped capillary reduces to a slit pore as $x\to\infty$. On
the other hand, the capping wall may provide sufficient energy for the fluid
inside the pore to form a liquid meniscus even when $P<P_{\text{v}}$. In the
latter case, as the pressure in the reservoir is increased towards CC ($P\to
P_{\text{v}}^{-}$), the meniscus should unbind continuously into the
``capillary bulk'' ($x\to\infty$), making CC a continuous (second-order)
phenomenon. The critical exponent for the diverging length of the liquid
slab, formed at the capping wall has been obtained analytically in reference
\cite{Parry07} and was confirmed numerically in reference
\cite{Yatsyshin2013}, for the case of dispersive fluid-fluid and
fluid-substrate interactions. An isotherm of drying (reservoir filled with
liquid, and $P\to P_{\text{v}}^{+}$) for a capillary with purely hard walls
and a fluid with short-ranged square well fluid-fluid potential is considered
by Roth and Parry in \cite{Roth11}.


Here we investigate the reasons behind the existence of two different regimes
of CC in a capped capillary pore: a continuous regime, when the filling is
happening ``from the surfaces'', and an abrupt one, when CC happens from
capillary bulk. The fluid-fluid and fluid-wall interactions are dispersive,
making the closest physical analogue Argon in contact with solid carbon
dioxide. We use a fully microscopic DF approach and provide density profiles,
adsorption isotherms and full phase diagrams.

\section{Free energy functional}
The equilibrium states of a statistical mechanical system correspond to the
minima of its free energy as a function of all parameters describing these
states (e.g., thermodynamic variables pressure, temperature and molar volume
for bulk fluids). In confined fluids, where structural inhomogeneities on a
micro-scale contribute significantly to the behaviour of the system, the
fluid structure itself constitutes such a ``parameter'' and has to enter the
expression for the free energy. Starting from a model for the molecular
interactions in the system, it is possible to construct an expression for the
free energy in the form of a functional of the one-body fluid density
$\rho\left({\bf r}\right)$. In the present study we utilise an application of
DF theories to the general case of confined atomic fluids, whose microscopic
structure is well described by generalized Van der Waals theories
\cite{ChandlerWeeksScience83,Hansen}.

When a substrate is brought in contact with a large reservoir containing the
fluid at some bulk pressure and temperature and the system is equilibrated,
the fluid density minimizes the grand free energy functional
$\Omega\left[\rho\left({\bi r}\right)\right]$:
\begin{equation}
    \label{Om}
    \Omega\left[\rho\left(\bi{r}\right)\right]=F_{\text{in}}\left[\rho\left(\bi{r}\right)\right]+ \int d\bi{r}\rho\left(\bi{r}\right)V\left({\bf r}\right)-\mu\int d\bi{r}\rho\left(\bi{r}\right),
\end{equation}
where $F_{\text{in}}\left[\rho\right]$ is the intrinsic free energy,
determined by the molecular interactions in a free fluid, $V\left({\bf
r}\right)$ is the effective potential, describing the effect of the substrate
upon the fluid (in our case the substrate is modelled as an inert spectator
phase, which cannot be affected by the fluid), and finally, $\mu$ is the
chemical potential of the thermostat, which together with $T$ (which enters
the expression for $F_{\text{in}}$) specifies the bulk thermodynamic fields
in the problem.

A comment is in order at this point, frequently omitted in the DF literature.
In a bulk fluid in isothermal conditions, due to the equation of state, the
pressure, $P=P_{\text{v}}$, is in one-to-one correspondence with the chemical
potential, $\mu$. Thus, the thermodynamic point of the bulk fluid at a given
value of $T$ can be specified by prescribing a value of either $P_{\text{v}}$
or $\mu$. However, when the fluid is brought in contact with the substrate,
it becomes inhomogeneous and the interfacial stress can no longer be
described by a scalar parameter ($P_{\text{v}}$), but requires defining a
space-dependant pressure tensor. The chemical potential, on the other hand,
is still constant throughout the fluid, as can be seen from equation
\eref{Om}, and together with the constant temperature enters the set of
thermodynamic fields describing an inhomogeneous fluid. Thus, the common in
DF models choice of $\mu$ over $P_{\text{v}}$ as the parameter of the
reservoir is more than just a convenience stemming from the form of equation
\eref{Om}.

\subsection{Model for the fluid}
In atomic fluids, molecular interactions at long ranges are well approximated
by London forces with a pairwise Lennard-Jones (LJ) potential \cite{whatliq}:
\begin{equation}
\label{LJ}
    \varphi^{6-12}_{\varepsilon_0,\sigma_0}\left(r\right) = 4\varepsilon_0\left[-\left(\frac{\sigma_0}{r}\right)^{6}+\left(\frac{\sigma_0}{r}\right)^{12}\right],
\end{equation}
where $\varepsilon_0$ and $\sigma_0$ are measures of the strength and range
of the potential, respectively. The short-ranged interactions and
correlations are dominated by repulsions, facilitating the use of a
perturbation scheme with the reference system being a fluid of purely
repulsive hard spheres, along with the random phase approximation for the
attractive direct pair-correlation function \cite{EvansSchool09}:
\begin{eqnarray}
    \label{Fin}
\fl F_{\text{in}}\left[\rho\left(\bi{r}\right)\right] = \int d \bi{r} \left(f_{\text{id}}\left(\rho\left({\bf r}\right)\right)\right)+F_{\text{hs}}\left[\rho\left({\bf r}\right)\right]\nonumber\\
    +\frac{1}{2}\int d \bi{r}\int d \bi{r}^{\prime}\rho\left(\bi{r}\right)\rho\left(\bi{r}^{\prime}\right)\varphi_\text{attr}\left(\left|{\bi{r}}-\bi{r}^{\prime}\right|\right),
\end{eqnarray}
where
$f_{\text{id}}\left(\rho\right)=k_{\text{B}}T\rho\left(\ln{\lambda\rho}-1\right)$
is the free energy of the ideal gas with the Boltzmann factor,
$k_{\text{B}}$, and the thermal wavelength, $\lambda$,
$F_{\text{hs}}\left[\rho\right]$ is the free energy of a purely repulsive
hard sphere fluid, and $\varphi_\text{attr}\left(r\right)$ is the effective
potential describing molecular attractions. Using the Barker and Henderson
prescription, we have \cite{BH}
\begin{equation}
\label{BH}
    \varphi_{\mbox{attr}}\left(r\right) = \left\{
    \begin{array}{lr}
        0, &r\leq \sigma\\
        \varphi^{\text{6-12}}_{\epsilon,\sigma},&r>\sigma
    \end{array}
    \right.,
\end{equation}
where $\epsilon$ is an effective depth of the attractive well and $\sigma$ is
the diameter of the reference hard sphere fluid.

For the hard sphere fluid we take the configurational part of the free energy
corresponding to the semi-phenomenological Carnahan-Starling equation of
state \cite{CS}:
\begin{eqnarray}
    \label{psi}
    \psi\left(\rho\right)= k_{\text{B}}T\frac{\eta\left(4-3\eta\right)}{\left(1-\eta\right)^2},\quad\eta=\pi\sigma^3\rho/6.
\end{eqnarray}
The hard sphere free energy in the second term of equation \eref{Fin} can then be written as
\begin{equation}
\label{Fhs}
F_{\text{hs}}\left[\rho\right] = \int d{\bf r}\rho\left({\bf r}\right)\psi\left(\bar{\rho}\left({\bf r}\right)\right),
\end{equation}
where $\bar{\rho}$ is a spatially averaged fluid density, which corresponds
to a weighted density approximation (WDA) and accounts for the non-local
correlations in the hard sphere fluid leading, e.g., to the oscillatory
near-substrate structure of the density profile. In this work we will use a
prescription for averaging due to Tarazona and Evans \cite{TarazonaWDA0},
which reproduces at distance $\sigma$ the step of the direct correlation
function of the hard sphere fluid, thus capturing the leading-order
excluded-volume effect \cite{TarazonaMelt}:
\begin{equation}
\label{WDArho}
    \bar{\rho}\left({\bf r}\right) =\frac{3}{4\pi\sigma^3} \int \limits_{\left|{\bf r}-{\bi r}^{\prime}\right|\leq\sigma} d{\bf r}^{\prime}\rho\left({\bf r}^{\prime}\right).
\end{equation}
Alternatively, one can use a local density approximation (LDA) \cite{TarazonaNonLoc}, by setting
\begin{equation}
\label{LDArho}
\bar{\rho}\left({\bf r}\right) \equiv {\rho}\left({\bf r}\right).
\end{equation}
Using LDA is less demanding computationally, which becomes important when
investigating numerically the parameter space of the model. Comparing the
results obtained from WDA and LDA one can understand the relative effect of
attractive molecular interactions with respect to the repulsive interactions.

\subsection{Model for the substrate}
We consider a physically realistic case, where the interactions between the
fluid and solid particles are of the same nature as the fluid-fluid
interactions, i.e. governed at long distances by the dispersive London forces
with the pairwise LJ potential, given in equation \eref{LJ}, with
substrate-specific parameters $\varepsilon_0=\varepsilon_{\text{w}}$,
$\sigma_0=\sigma_{\text{w}}$. It is known that in wetting (unlike dewetting)
the end-effects are negligible \cite{JChemPhys92}, which allows us to model
the wetting on a capped capillary of a large length by considering a
semi-infinite pore, figure \ref{FigOne}. As is typical in the DF theories of
fluids with vapour- and liquid-like densities, we model the solid substrate
as an inert spectator phase, accounting for its presence by an additive
external potential (second term in equation \eref{Om}). A different model for
the substrate, where it, in turn, can be affected by the fluid, has been
adopted by Ustinov and Do in \cite{DDDo05} for wetting on a cylinder.

The external potential due to the substrate acting on the fluid is obtained
by integrating the LJ potential over the volume of the substrate. For the
capped capillary we assume that three semi-infinite pieces of material have
been brought together in a manner shown in figure \ref{FigOne}. Since the
intermolecular potential at short distances is very different from LJ
\cite{whatliq, IspolatovWidomPhysicaA00}, we introduce a cutoff,
$H_0\geq\sigma^{\left(\text{i}\right)}$, separating the substrate from the
fluid. This has the added numerical benefit of avoiding the divergence of the
external potential at short distances, which leads to a super-exponential
decay of the fluid density near the walls and requires special computational
tricks to be resolved \cite{Yatsyshin2012}. Experimentally a cutoff can be
implemented by coating the substrate with a layer of a foreign species
\cite{AncilottoToigoJChemPhys00}.

By integrating the pairwise LJ potential, equation \eref{LJ}, over the volume
of the substrate of constant density $\rho_{\text{w}}$, we arrive at the
expression for the potential $V\left({\bf r}\right)\equiv V\left(x,y\right)$
exerted on the fluid by the three walls of the capillary (see chunks 1 -- 3
in figure \ref{FigOne}):

\begin{eqnarray}
    \label{Vxy}
    \fl V\left(x,y\right)=\rho_{\text{w}}\int_{-\infty}^{\infty}dz^{\prime}\times   \left(\int_{-\infty}^{\infty}dx^{\prime}\times\left(\int_{-\infty}^{-H_0}dy^{\prime} + \int_{H_0}^{+\infty}dy^{\prime}\right)+\int_{-\infty}^{-H_0}dx^{\prime}\int_{-H_0}^{H_0}dy^{\prime}\right)\nonumber\\
                   \times\varphi^{\text{6-12}}_{\varepsilon_{\text{w}},\sigma_{\text{w}}}\left(\sqrt{\left(x-x^{\prime}\right)^2+\left(y-y^{\prime}\right)^2+{z^{\prime}}^2}\right).
\end{eqnarray}

\subsection{Comments}
The approximations giving rise to our free energy functional (equations
\eref{Om}, together with \eref{Fin} -- \eref{Vxy}) follow the generalized Van
der Waals picture of atomic fluids, which is supported by existing
observations \cite{Plischke,Hansen}. The model is able to capture the
dominant physical effects determining the qualitative behaviour of the system
in the range of liquid-like and vapour-like fluid densities. However,
obtaining quantitative information for a particular system would require a
more detailed model.

For example, there exist more sophisticated hard sphere functionals, e.g.
\cite{Tar97,WhiteBear}, which accurately describe the full repulsive direct
correlation function for a wide range of temperatures. However, the excluded
volume effects do not play a significant role in the process of wetting
\cite{Yatsyshin2012,Evans90, TarazonaNonLoc}, as is also demonstrated with
calculations in the following sections.

The attractive functional (last term in equation \eref{Fin}) captures the
asymptotic tail of the ``attractive'' contribution to the direct correlation
function of a bulk fluid, which along with a non-local treatment of the
density, suffices to capture the physics of wetting by vapour or liquid
\cite{Hansen}. In a more complete treatment, the hard sphere diameter,
$\sigma$, should be made weakly temperature-dependent, but in the absence of
a unified approach, the choice of a particular prescription depends on the
range of temperatures
\cite{WaltonQuirkeChemPhysLett86,CottermanetalAlChE86,LuetalMolPhys85}. The
attractive functional can have the same non-local form as the repulsive one
(equation \eref{Fhs}), with the analogues of $\psi\left(\bar{\rho}\right)$
and the averaging procedure for $\bar{\rho}$ obtained using perturbation
theory around a reference uniform bulk fluid, e.g. \cite{LeeEtAl76}, or by
functional integration, starting from a reasonable analytic approximation to
the bulk attractive direct correlation function
\cite{Zeng10,Swe01,ThieleJChemPhys63}. Finally, the three-body attractive
interactions can also contribute to the values of quantities that can be
observed experimentally, such as adsorption, e.g. reference
\cite{BarkerEtAlMolPhys71}.

An optimal model for the substrate potential can use \emph{ab initio} results
\cite{Chiz98,Marsh96}, and possibly account for the effect of the fluid on
the substrate, instead of modelling it as an inert phase \cite{DDDo05}.

In the apparent absence of a generally accepted approach, which would allow
to retrieve observed values and resolve the specific peculiarities of a given
system, the main value of DF models is, first, in their theoretical
consistency, which allows us to use the inter-particle potentials as input
and, by incrementally including finer effects, to uncover the dominant
physical processes contributing to the observed qualitative behaviour.
Second, a detailed parametric investigation of DF models allows us to uncover
new phenomena, which may not yet be accessible experimentally (as was the
case with first- and second- order wall wetting during their discovery in
reference \cite{EbnerSaam}). Finally, DF investigations serve as guides to
both experimentalists and theoreticians (e.g., in the development of
analytical approaches, like effective Hamiltonians).

\section{Working equations}
Given the reservoir parameters $T$ and $\mu$, the Euler-Lagrange equation for
the minimization of $\Omega\left[\rho\right]$ defined in equation \eref{Om},
has the form
\begin{eqnarray}
\label{EL3D}
\fl T\ln{\rho}\left({\bf r}\right)+\psi\left(\rho\left({\bf r}\right)\right)+
\int d{\bf r^\prime}\rho\left({\bf r^\prime}\right)\psi_{\rho}^\prime\left(\bar{\rho}\left({\bf r^\prime}\right)\right) W\left({\bf r}-{\bf r^\prime}\right)\nonumber\\
+\int d{\bf r^\prime}\rho\left({\bf r^\prime}\right)\varphi_{\text{attr}}\left(\left|{\bf r}-{\bf r^\prime}\right|\right)
+V\left({\bf r}\right)-\mu=0,
\end{eqnarray}
where the integration is carried out over the entire volume occupied by the
fluid ($0\leq x\leq\infty$, $0\leq y\leq H$, figure \ref{FigOne}),
$\psi^\prime_{\rho}$ is the derivative of the configurational hard sphere
free energy (equation \eref{psi}) with respect to the density and
$W\left({\bf r}\right)$ is the weight function defining $\bar{\rho}$ in
equations \eref{WDArho}, \eref{LDArho} through volume integrals with step-
and delta-functions:
\begin{equation}
\label{weight3D}
    W\left({\bf r}\right) = \left\{
    \begin{array}{lr}
          \displaystyle \frac{3}{4\pi\sigma^3}\enskip\Theta\left(\sigma-r\right)&\text{for WDA},\vspace{1em} \\
         \displaystyle\delta\left({\bf r}\right)&\text{for LDA}.
    \end{array}
    \right.
\end{equation}

When the density of the fluid is constant in one or more directions, the
integration along those directions can be carried out analytically in
equation \eref{EL3D}, reducing its dimensionality by simplifying the
expressions for $\varphi_{\text{attr}}$ (equation \eref{BH}) and $W$
(equation \eref{weight3D}). To study the phase behaviour of the fluid in the
capped capillary, we need to be able to find the densities in the capillary
and in the slit pore, to which it reduces to as $x\to\infty$ (thus the slit
pore forms the capillary bulk). The former corresponds to a two-dimensional,
and the latter -- to a one-dimensional (1D) reduction of the general equation
\eref{EL3D}. The expressions for both geometries are given in the following
subsections. From now on we set the parameters of fluid-fluid interactions
$\varepsilon$ and $\sigma$ to be units of energy and length.

\subsection{1D problem}
For the 1D problem of the slit pore immersed in vapour we have in equation
\eref{EL3D}: ${\bf r}\equiv y\cdot{\bf e_y}$, $\rho\left({\bf
r}\right)\equiv\rho^{\text{slt}}\left(y\right)$. The substrate potential,
$V\left({\bf r}\right)\equiv V^{\text{slt}}\left(y\right)$, is defined as the
limit of the expression in equation \eref{Vxy} as $x\to\infty$
\cite{Yatsyshin2012}:
\begin{eqnarray}
    \label{Vy}
    V^{\text{slt}}\left(y\right) = V^{3-9}_{\varepsilon_{\text{w}},\sigma_{\text{w}}}\left(y\right)+V^{3-9}_{\varepsilon_{\text{w}},\sigma_{\text{w}}}\left(H-y\right), \nonumber\\
    V^{3-9}_{\varepsilon_{\text{w}},\sigma_{\text{w}}}\left(x\right)\equiv 4\pi\varepsilon_{\text{w}}\rho_{\text{w}}\sigma_{\text{w}}^3\left(-\frac{1}{6}\left(\frac{\sigma_{\text{w}}}{x}\right)^3+\frac{1}{45}\left(\frac{\sigma_{\text{w}}}{x}\right)^9\right).
\end{eqnarray}
The attractive potential, $\varphi_{\text{attr}}\left(y\right)\equiv\int dz \int dx\enskip
\varphi_{\text{attr}}\left(\sqrt{x^2+y^2+z^2}\right)$, is obtained by integrating equation \eref{BH}:
\begin{equation}
    \label{phi1D}
    \varphi_{\text{attr}}\left(y\right) = \left\{
            \begin{array}{lr}
                                               \displaystyle -\frac{6\pi}{5}, & \text{ if }\left|y\right|\leq1, \vspace{1em}\\

               \displaystyle 4\pi\left(\frac{1}{5y^{10}}-\frac{1}{2y^4}\right), & \text{ if }\left|y\right|>1.
            \end{array}
            \right.
\end{equation}
The weight function, $W\left(y\right)\equiv\int dz \int dx\enskip
W\left({x\cdot{\bf e_x}+y\cdot{\bf e_y}+z\cdot{\bf e_z}}\right)$, is obtained by integrating equation \eref{weight3D}:
\begin{equation}
\label{weight1D}
    W\left(y\right) = \left\{
    \begin{array}{lr}
          \displaystyle \frac{3}{4}\left(1-y^2\right)\Theta\left(1-y\right)&\text{for WDA},\vspace{1em} \\
         \displaystyle\delta\left(y\right)&\text{for LDA}.
    \end{array}
    \right.
\end{equation}

\subsection{2D problem}
For the 2D problem of the capped capillary immersed in vapour we have in
equation \eref{EL3D}: ${\bf r}\equiv x\cdot{\bf e_x}+y\cdot{\bf e_y}$,
$\rho\left({\bf r}\right)\equiv\rho^{\text{cpd}}\left(x,y\right)$ and the
substrate potential $V\left({\bf r}\right)\equiv
V^{\text{cpd}}\left(x,y\right)$ is defined in equation \eref{Vxy}. The 2D
reduction of equation \eref{EL3D} can be done analytically: the integrals
over $z$ can be obtained in a closed form. Unfortunately, some of the
resulting expressions are very complicated and lead to significant rounding
errors if used in computations. For that reason we use a hybrid approach and
evaluate some of the $z$-integrals in equation \eref{EL3D} numerically.

First, let us integrate out $z$ in the pairwise LJ potential in equation
\eref{LJ}, and formally define an expression that resembles the 2D case where
the fluid density is constant in one direction,
$\varphi^{5-11}_{\varepsilon_0,\sigma_0}\left(r\right)\equiv\int dz\enskip
\varphi^{\text{6-12}}_{\varepsilon_0,\sigma_0}\left(\sqrt{x^2+y^2+z^2}\right)$:
\begin{eqnarray}
\label{LJxy}
\varphi^{5-11}_{\varepsilon_0,\sigma_0}\left(r\right) = \frac{3\pi\varepsilon_0\sigma_0}{2}\left[-\left(\frac{\sigma_0}{r}\right)^{5}+\frac{21}{32}\left(\frac{\sigma_0}{r}\right)^{11}\right],
\end{eqnarray}
where (as everywhere for the 2D problem) $r=\sqrt{x^2+y^2}$.

Now the expression for $V\left(x,y\right)$ can be written in the form:
\begin{equation}
V^{\text{cpd}}\left(x,y\right) = V^{\text{slt}}\left(y\right)+V^{\text{cap}}\left(x,y\right),
\end{equation}
where $V^{\text{slt}}\left(y\right)$ is the potential of a slit pore
(equation \eref{Vy}) and $V^{\text{cap}}\left(x,y\right)$ accounts for the
contribution due to the capping wall and is determined by integrating the
fluid-substrate potential,
$\varphi^{5-11}_{\varepsilon_{\text{w}},\sigma_{\text{w}}}$ (see equation
\eref{LJxy}), over chunk 2 in figure \ref{FigOne}:
\begin{equation}
\label{Vcap}
V^{\text{cap}}\left(x,y\right) = \int\limits_{-\infty}^{-H_0}dx^\prime\int\limits_{-H_0}^{H+H_0}dy^\prime\enskip\varphi^{5-11}_{\varepsilon_{\text{w}},\sigma_{\text{w}}}\left(\left|{\bf r}-{\bf r^\prime}\right|\right),
\end{equation}
Any further analytic simplification of equation \eref{Vcap} leads to high
numerical instabilities in the resulting expression, so we have preferred to
compute $V^{\text{cap}}\left(x,y\right)$ in equation \eref{Vcap} numerically,
using the Clenshaw-Curtis quadrature \cite{Yatsyshin2012}.

Consider now the attractive potential defined in the 2D problem as
$\varphi_{\text{attr}}\left(r\right)\equiv\int dz\enskip
\varphi_{\text{attr}}\left(\sqrt{x^2+y^2+z^2}\right)$. Integrating out
$z$-coordinate in equation \eref{BH}, we obtain
\begin{equation}
    \label{phi2D}
    \varphi_{\text{attr}}\left(x,y\right) = \left\{
            \begin{array}{lr}
                                               \displaystyle 2\int\limits_{\sqrt{1-r^2}}^{\infty}dz\enskip\varphi^{6-12}_{1,1}\left(\sqrt{r^2+z^2}\right), &\text{if }r\leq1, \vspace{1em}\\

               \displaystyle \varphi^{5-11}_{1,1}\left(r\right), &\text{ if }r>1.
            \end{array}
            \right.
\end{equation}

Noteworthy is that the integral in \eref{phi2D} can be done analytically
\cite{PereiraKalliadasisJFM12}, but the resulting expression is again
unstable numerically, leading to high rounding errors, when used in actual
calculations. So a numerical quadrature is preferable.

Finally, the 2D weight function, $W\left(x,y\right)\equiv\int dz\enskip
W\left({x\cdot{\bf e_x}+y\cdot{\bf e_y}+z\cdot{\bf e_z}}\right)$, is obtained from equation \eref{weight3D}:
\begin{equation}
\label{weight2D}
    W\left(x,y\right) = \left\{
    \begin{array}{lr}
          \displaystyle \frac{3\pi}{2}\left(1-r^2\right)\Theta\left(1-r\right)&\text{for WDA},\vspace{1em} \\
         \displaystyle\delta\left({\bf r}\right)&\text{for LDA}.
    \end{array}
    \right.
\end{equation}

Note that although we have defined the 2D pairwise potential,
$\varphi^{5-11}_{\varepsilon_0,\sigma_0}$ (equation \eref{LJxy}), its use in
our expressions is justified by formally changing the orders of integration
and reflects a simple fact that the fluid density is constant along direction
${\bf e_z}$. The fluid dimensionality remains identical to the
three-dimensional hard sphere fluid described by the Carnahan-Starling
equation of state (see equation \eref{psi}).

\subsection{Numerical strategy}
The numerical approaches to solving equation \eref{EL3D} can be categorized,
first, with respect to the method for evaluating the non-local integrals and
second, with respect to the iterative method for solving the discretized
equation \cite{FinkSal3}.

The convolution-like form of non-local terms prompts many authors to advocate
a fast Fourier transform. Recent works include, e.g., references
\cite{KnepleyEtAlJChemPhys10, Roth10}. The shortcomings are, first, that such
methods often require equidistant grid points, which is not necessary in the
regions where density is nearly constant. Second, Fourier-based methods on
non-periodic domains are equivalent to a higher order Simpson quadrature and
possess an algebraic convergence rate. Finally, such methods often impose
non-physical periodicities of solution. We believe that the use of a
specialized quadrature in real space is preferable. A combination of a Gauss
quadrature and an algebraically converging trapezoid rule was proposed in
reference \cite{HS3D}.

In the present work we use a novel numerical approach based on the Chebyshev
pseudo spectral collocation method \cite{Boyd, Trefethen}. The problem is
descretized on a non-uniform grid of collocation points, with the solution
obtained in the form of a rational interpolant \cite{TT}. Using conformal
maps from a unitary circle in the complex plane, we can control the
distribution of the collocation points on the calculation domain
concentrating them in the regions, where the solution exhibits steep
gradients (near walls, near liquid -- vapour interface) and using a moderate
number of points in the regions of regularly behaving or near-constant
density (inside liquid/vapour phases). We evaluate all integral terms in real
space by a highly accurate Clenshaw-Curtis quadrature \cite{HT,Wald03}, which
in the latest numerical analysis literature has been found to be an optimal
choice with respect to convergence rates, accuracy and stability
\cite{Tref08}. Our numerical method possesses an exponential convergence
rate, while at the same time allowing to use a moderate number of collocation
points. The interested reader is referred to our earlier work
\cite{Yatsyshin2012}, where all the details of implementation, along with
convergence tests and examples are provided.

In the present study we extend the method used in reference
\cite{Yatsyshin2012}, which could only deal with 1D problems, by utilising
the appropriate interpolant on a tensor product grid and constructing
integration matrices for 2D geometries that can be mapped to a square. Even
though the computations necessary to construct the integration matrices
become rather involved, a significant advantage is that such calculation
needs to be done only once and the matrix can be stored and reused for other
calculations in the same geometry, such as for obtaining isotherms and phase
diagrams. Moreover, the aforementioned exponential accuracy is even more
important in a 2D geometry, as it allows us to obtain very accurate results
with only a moderate number of grid points


With respect to an iterative procedure for the spatially discretized problem,
the numerical methods for DF calculations mostly use self-consistent Picard
iterations. The tradition of that approach goes back to the earliest works in
the 1970s. The method is highly numerically unstable, often requires \emph{ad
hoc} modifications of consecutive iterations \cite{Roth10}. Furthermore, the
convergence is typically achieved after hundreds or thousands of iterations,
and in the case of multiple existing solutions (e.g., when the system
undergoes a first-order phase transition) further \emph{ad hoc} modifications
drawing heavily on one's physical intuition might be necessary to customize
the numerical method for a particular system, e.g.
\cite{KnepleyEtAlJChemPhys10}.

We completely automate the numerical solution of DF equations for systems
with and without phase transitions using a Newton iteration procedure in
conjunction with an arc-length continuation technique \cite{Yatsyshin2012}.
Such approach was first used for DF calculations by Salinger and Frink in
\cite{FrinkSalinger1}. In our implementation the convergence is typically
achieved in 2 -- 3 iteration steps, and we believe the Newton solver or its
variation to be an optimal strategy for 1D and 2D problems. However, unlike
Piccard, the Newton method requires to find and invert the Jacobian matrix on
every iteration, which may shift optimality back to a variation of a
self-consistent approach (they typically do not require expensive matrix
inversions) in three-dimensional problems \cite{FinkSal3,HS3D}.

In general, the arc-length continuation technique can be used to obtain sets
of solutions to equation \eref{EL3D} corresponding to a variable parameter
(typically, a thermodynamic field acting in the system). By treating the
parameter as an unknown and supplementing \eref{EL3D} the procedure with a
geometrical constraint in the space of the discretized problem, one can
systematically obtain all the possible solution-parameter combinations
consistent with the equation \eref{EL3D}. For example, choosing the chemical
potential (or the bulk density) as the variable parameter, allows us to
obtain adsorption isotherms \cite{FrinkSalinger1}, while choosing temperature
would produce an isochore, e.g. \cite{TeloDaGamaMarconiPhysicaA91}. One can
also study the dependence of density on the parameters of the fluid-fluid or
fluid-wall potentials \cite{FrinkSalinger1,Yatsyshin2012}.

\subsection{Isotherms and phase diagrams}
A single density profile carries no information about the thermodynamic
stability of the fluid state, since a particular solution of equation
\eref{EL3D} might not minimise $\Omega\left[\rho\left({\bf r}\right)\right]$,
but can correspond to its saddle point in the phase space. So finding a
density profile by solving equation \eref{EL3D} does not guarantee, that the
corresponding fluid state is physical. On the other hand, computing a family
of solutions parametrised by a thermodynamic field allows to determine the
stability of fluid configurations by finding $\Omega\left[\rho\right]$ for
each of the calculated density profiles. The resulting dependence of the
fluid free energy on the thermodynamic field is a concave function, when the
fluid is in a stable equilibrium \cite{GriffithsWheelerPhysRevA70}. We will
be using such an approach extensively in this work~\cite{HendersonBook92}.
Apart from determining the thermodynamic stability of fluid states it allows
to locate phase transitions.

For a fluid configuration inside the capped capillary we define the free energy excessive over capillary bulk as
\begin{equation}
\label{OmEx}
\Omega^{\text{ex}}=\Omega\left[\rho^{\text{cpd}}\left(x,y\right)\right]-\Omega\left[\rho^{\text{slt}}\left(y\right)\right],
\end{equation}
where the value, given by equation \eref{Om}, is computed for the solutions
to \eref{EL3D} on the functional spaces of 2D and 1D density profiles,
respectively. The excess free energy is thus a function of thermodynamic
fields acting in the system and carries information about effects specific
only to the 2D confinement of the fluid. Thermodynamically stable fluid
states belong to its concave branches in the space of all fields acting in
the system \cite{HendersonBook92}. In order to reveal the phase behaviour of
the fluid we obtain characteristic level sets of $\Omega^{\text{ex}}$ -- the
isotherms.

The set of thermodynamic fields in our problem consists of the parameters of
the pairwise fluid-substrate potential, the width of the capillary and the
bulk fields $T$ and $\mu$. While we will reveal the effects of all relevant
fields on the phase behaviour of the system, for convenience and consistency
we will be considering various isothermal thermodynamic routes at fixed
values of substrate parameters. Thus, the thermodynamic variable allowed to
change is $\mu$. Note that in a macroscopic description by equation
\eref{Kelvin}, $P_{\text{l}}$ and $P_{\text{v}}$ are in one-to-one
correspondence with $\mu$, due to the bulk equation of state.

The thermodynamic density conjugate to $\mu$ is adsorption, $\Gamma$, related to $\Omega^{\text{ex}}$ through the Gibbs equation \cite{HendersonBook92}:
\begin{equation}
\label{Gibs}
\Gamma=-\partial\Omega^{\text{ex}}/\partial\mu,
\end{equation}
where the derivative is evaluated at a fixed value of $T$. In our study
$\Gamma$ acts as a natural \emph{order parameter}. A first-order surface
phase transition is associated with a finite jump in $\Gamma$, and a
continuous surface phase transition -- with the divergence of $\Gamma$ as a
function of one of the thermodynamic fields. At any given value of $T$ we
count the value of $\mu$ from bulk saturation,
$\mu_{\text{sat}}\left(T\right)$, thus the \emph{control parameter} is the
disjoining chemical potential, $\Delta\mu$, defined as
$\Delta\mu=\mu-\mu_{\text{sat}}$.

For the grand canonical ensemble in isothermal conditions there exists an
exact sum rule, which relates adsorption to the fluid one-body density
profile \cite{Swol86}:
\begin{equation}
\label{Gam}
\Gamma=\int{dx\ dy}\left(\rho^{\text{cpd}}\left(x,y\right)-\rho^{\text{slt}}\left(y\right)\right).
\end{equation}
The above equation also suggests, that $\Gamma$ is a convenient measure for
the amount of ``excess over bulk'' fluid, adsorbed on the substrate surface
(in our case -- the capping wall and corners), hence the commonly used term
-- adsorption.

By using arc-length continuation for equation \eref{EL3D}, treating $\mu$ as
the continuation parameter, we compute sets of points on an isotherm
$\Omega^{\text{ex}}\left(\Delta\mu\right)$. Each data point on the isotherm
corresponds to a density profile. Computing free energy isotherms allows us
to identify thermodynamically stable density profiles, as they correspond to
the data points on the concave branches of
$\Omega^{\text{ex}}\left(\Delta\mu\right)$. It also allows us to detect phase
transitions: a first-order transition is manifested by a Van der Waals loop
(the intersection of concave branches), and a continuous phase transition --
by a finite limit of the free energy isotherm and a divergence of the
adsorption isotherm. For every value of $T$ we additionally make sure that
our calculation of the isotherms $\Omega^{\text{ex}}\left(\Delta\mu\right)$
and $\Gamma\left(\Delta\mu\right)$ is correct by computing
$\Gamma\left(\Delta\mu\right)$ from equation \eref{Gam} and comparing it
versus the values obtained independently from the Gibbs equation \eref{Gibs},
which relates adsorption and free energy isotherms.

Given a set of points on an isotherm exhibiting a Van der Waals loop, we find
the exact location of the first-order phase transition in the $\Omega$ --
$\mu$ plane and the coexisting density profiles, $\rho_1$ and $\rho_2$, by
solving a system of three equations: the two Euler-Lagrange equations for
$\rho_1$ and $\rho_2$, supplemented with the condition of equal grand free
energies:
\begin{eqnarray}
\label{tr}
\frac{\delta\Omega}{\delta\rho}\Big |_{\rho_1}=\frac{\delta\Omega}{\delta\rho}\Big |_{\rho_2}=0\nonumber\\
\Omega^{\text{ex}}\left[\rho_1\right]-\Omega^{\text{ex}}\left[\rho_2\right]=0,
\end{eqnarray}
where for the capped capillary we use the excess free energy
$\Omega^{\text{ex}}\left[\rho\right]\equiv\Omega^{\text{ex}}\left[\rho^{\text{cpd}}\left(x,y\right)\right]$
defined in equation \eref{OmEx}, and for the slit pore (capillary bulk) --
the full grand free energy given by equation \eref{Om}, i.e
$\Omega^{\text{ex}}\left[\rho\right]\equiv\Omega\left[\rho^{\text{slt}}\left(y\right)\right]$.
We use the data from the isotherm to select an initial guess for solving
equation \eref{tr}. Normally we chose the two density profiles, whose
corresponding free energies belong to the intersecting branches of
$\Omega^{\text{ex}}\left(\Delta\mu\right)$ and lie close to the intersection
point in the $\Omega$ -- $\mu$ plane.

By applying the arc-length continuation to the system of equations \eref{tr}
and treating $T$ as the continuation parameter we can obtain the phase
diagram in the $T$ -- $\mu$ plane, thus completely describing the
thermodynamic phase behaviour of the system. In what follows we discuss the
calculations of isotherms and full phase diagrams in detail in the following
section.

\section{Physics of wetting on a capped capillary}

In this section we present a number of representative calculations taken from
a detailed numerical study of the parameter space and discuss the possible
adsorption scenarios and surface phase transitions occurring in the system.
More specifically, we consider several examples of capped capillaries with
various substrate parameters and argue in terms of complete phase diagrams in
the $T$ -- $\Delta\mu$ space together with wetting isotherms of excess free
energy, $\Omega^{\text{ex}}\left(\Delta\mu\right)$, and adsorption,
$\Gamma\left(\Delta\mu\right)$. The disjoining chemical potential,
$\Delta\mu$, is given as the difference of the applied chemical potential,
$\mu$, from the saturation chemical potential at bulk liquid -- vapour
coexistence,  $\mu_{\text{sat}}$.

We note that the choice of particular values for parameters in the governing
equations was affected (without loss of generality) by numerical convenience.
For example, we tried to select substrate potentials with relatively high
planar wetting temperatures, $T_{\text{w}}$, to avoid layering transitions
\cite{Dietrich, BallEv}, and also for the liquid -- vapour interfaces to be
sufficiently smooth and resolvable on a moderately dense mesh. Obviously, the
generality of our conclusions is not impaired by such considerations.

\subsection{Capillary prewetting}
\label{secLV} Consider a capillary with width $H=30$ and a substrate defined
by $\varepsilon_{\text{w}}=0.85$, $\sigma_{\text{w}}=1.35$ and $H_0=2.2$. The
fluid is treated within WDA. For the case of a planar wall in contact with
vapour we find the wetting temperature to be $T_{\text{w}} = 0.927$. We start
the investigation of the capped capillary by setting $T=0.93$, a more or less
arbitrarily chosen value in the region where a planar substrate wall exhibits
prewetting. We further set the chemical potential,
$\mu=\mu_{\text{sat}}+\Delta\mu$, in equation \eref{EL3D} to a rather low
value ($\Delta\mu=-5$), so that the fluid is well inside the bulk vapour
phase, and thus almost unaffected by the presence of the substrate. This
facilitates the convergence of Newton's scheme, where the initial guess is
taken to be that of the vapour inside a slit pore, at the same chemical
potential and temperature:
$\rho_{N_{xy}}^{\text{cpd},0}\left(x,y\right)\equiv\rho^{\text{slt}}\left(y\right)$.
After obtaining the solution $\rho^{\text{cpd}}_{N_{xy}}\left(x,y\right)$,
given on the spatial grid by the set of data points $\rho_{ij}$, we perturb
the vector $\left[\mu,\rho_{ij}\right]$ and, adding to the equation
\eref{EL3D} a geometric constraint of continuity in the
$\left(N_{xy}+1\right)$-space, we solve the resulting system of equations for
the next density profile and the corresponding $\mu$, also by the Newton
method. The described tactics is the essence of the arc-length continuation
technique, which allows us to systematically obtain consecutive solutions to
\eref{EL3D} at values of $\Delta\mu$ increasing towards CC, i.e. a wetting
isotherm. The details of implementation, along with a discussion of various
geometrical constraints, are covered in our previous work
\cite{Yatsyshin2012}. The main practical benefit is that the algorithm serves
to adjust each obtained solution to provide an optimal initial guess for
finding the next one, at a different value of the parameter ($\mu$). The
continuation must be initiated from a starting point -- a solution at some
known value of the parameter, which is why we start from a simple vapour
profile far from coexistence.

\begin{figure}
\centering
    \includegraphics{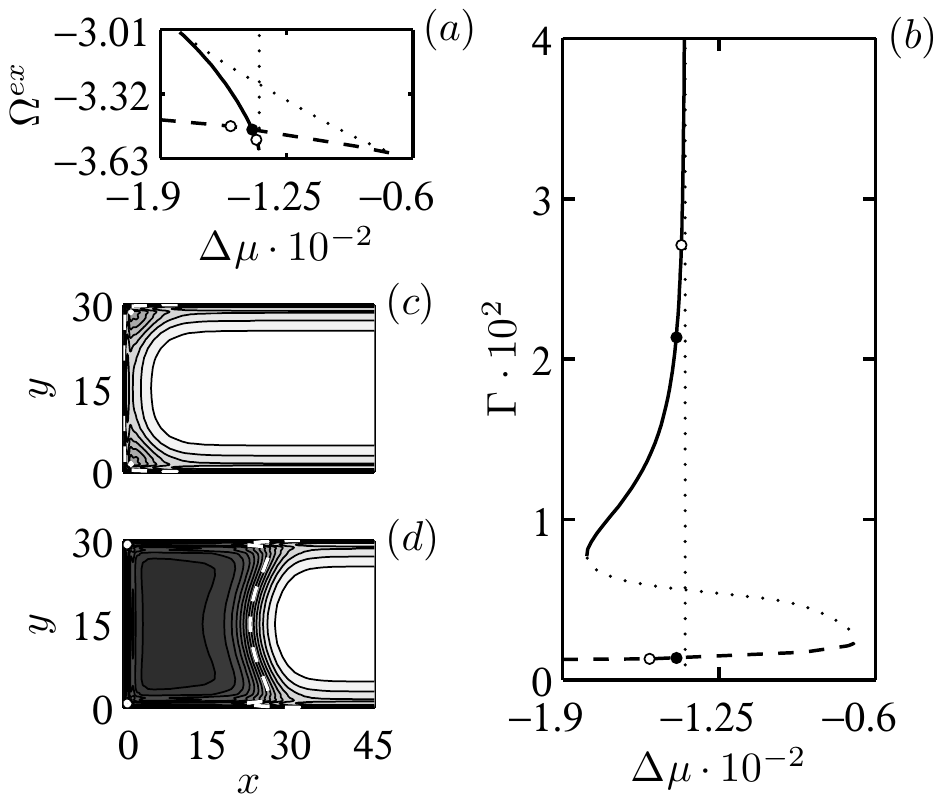}%
    \caption{Capillary prewetting transition at $T=0.93$, $\Delta\mu_{\text{cpw}}\left(T\right)=-1.43\cdot10^{-2}$ in the capped
    capillary with $H=30$, $\varepsilon_{\text{w}}=0.85$, $\sigma_{\text{w}}=1.35$, $H_0=2.2$;
    fluid treated in WDA, planar $T_{\text{w}} = 0.927$. CC of the associated slit pore (capillary bulk) is at
    $\Delta\mu_{\text{cc}}\left(T\right)=-1.39\cdot10^{-2}$. (a) Excess free energy isotherm. It has two
    concave branches connected by a non-concave branch (dotted line). The concave branches define two thermodynamically stable phases,
    coexisting at $\Delta\mu_{\text{pw}}$: vapour (dashed line, branch extends from $\Delta\mu=-\infty$, up to
    its spinodal at $\Delta\mu=-0.68\cdot10^{-2}$) and capillary-liquid slab (solid line, branch extends from its
    spinodal at $\Delta\mu=-1.80\cdot10^{-2}$ up to $\Delta\mu=\Delta\mu_{\text{cc}}$, indicated by vertical dotted line).
    Capillary prewetting is marked by the filled circle at $\Delta\mu_{\text{cpw}}$ and
    $\Omega^{\text{ex}}\left(\Delta\mu_{\text{cpw}}\right)=-3.49$. Open circles show the two continuation data points,
    whose corresponding density profiles were used as initial guess in equation \eref{tr}. (b)
    Adsorption isotherm. Line styles and open circles are defined as in plot (a). Note that
    $\Delta\mu=\Delta\mu_{\text{cc}}$ is the vertical asymptote for $\Gamma\left(\Delta\mu\right)$.
    Capillary prewetting corresponds to the jump of adsorption from $\Gamma_1=14$ to $\Gamma_2=214$, marked by filled circles.
    (c), (d) Coexisting density profiles. Data rescaled between $\rho_{\text{cc}}^{\text{vap}}\left(T\right)=0.1$ (white),
    and $\rho_{\text{cc}}^{\text{liq}}\left(T\right)=0.43$ (dark grey). The white dashed line indicates interface between
    vapour and capillary-liquid.
    \label{Figis1}}%
\end{figure}

We observe that at higher values of $\Delta\mu$ a liquid-like slab starts to
form near the capping wall. As the chemical potential approaches its value at
CC, $\Delta\mu_{\text{cc}}\left(T\right)$, the length of the slab increases
and diverges in the limit $\Delta\mu\to\Delta\mu_{\text{cc}}$. Examination of
the fluid structure (details are given in section \ref{StongerConfinement})
reveals, that the liquid-like phase forming the slab is identical to
capillary-liquid, while the vapour-like phase filling the capillary bulk is
identical to the vapour, where both the vapour and capillary-liquid are taken
at CC. The behaviour of the fluid is best understood in terms of isotherms
shown in figure \ref{Figis1}. A typical Van der Waals loop in the excess free
energy dependence, $\Omega^{\text{ex}}\left(\Delta\mu\right)$, in figure
\ref{Figis1}(a) indicates that a first-order phase transition occurs at
$\Delta\mu=\Delta\mu_{\text{cpw}}\left(T\right)$ (marked by a filled circle),
where the concave branches of excess free energy cross. The adsorption
isotherm, $\Gamma\left(\Delta\mu\right)$, presented in figure \ref{Figis1}(b)
possesses a  characteristic hysteresis behaviour, with the equal area
construction giving the same value for $\Delta\mu_{\text{cpw}}$, as the
intersection of branches of $\Omega^{\text{ex}}\left(\Delta\mu\right)$, thus
providing a test for our numerical implementation via an exact Gibbs phase
rule, equation \eref{Gibs}. The adsorption diverges as
$\Delta\mu\to\Delta\mu_{\text{cc}}\left(T\right)$ indicating that CC in a
capped capillary is a continuous phenomenon. The critical exponents for the
diverging height of the liquid slab were obtained analytically by Parry
\emph{et al} in \cite{Parry07} and numerically in our previous work in
\cite{Yatsyshin2013}.

The coexisting density profiles are shown in figures \ref{Figis1}(c) and
\ref{Figis1}(d). For illustration purposes, throughout the manuscript filled
contours of 2D density profiles are coloured in shades of gray between bulk
vapour-like and bulk liquid-like densities at CC, at the given temperature,
$\rho_{\text{cc}}^{\text{vap}}\equiv\rho_{\text{cc}}^{\text{vap}}\left(\Delta\mu_{\text{cc}}\left(T\right),T\right)$
(white) and
$\rho_{\text{cc}}^{\text{liq}}\equiv\rho_{\text{cc}}^{\text{liq}}\left(\Delta\mu_{\text{cc}}\left(T\right),T\right)$
(dark grey), respectively. The values of $\rho_{\text{cc}}^{\text{vap}}$ and
$\rho_{\text{cc}}^{\text{liq}}$ are obtained by assuming a uniform density
distribution and solving equation  \eref{EL3D} in the absence of the wall
potential ($V\left({\bf r}\right)\equiv0$, $\rho\left({\bf
r}\right)\equiv\rho=\text{const}$). We define the sharp liquid -- vapour
interface (dashed white line) by the Gibbs dividing surface, forming the
level set at the value
$\left(\rho_{\text{cc}}^{\text{vap}}+\rho_{\text{cc}}^{\text{liq}}\right)/2$.
\begin{figure}
\centering
    \includegraphics{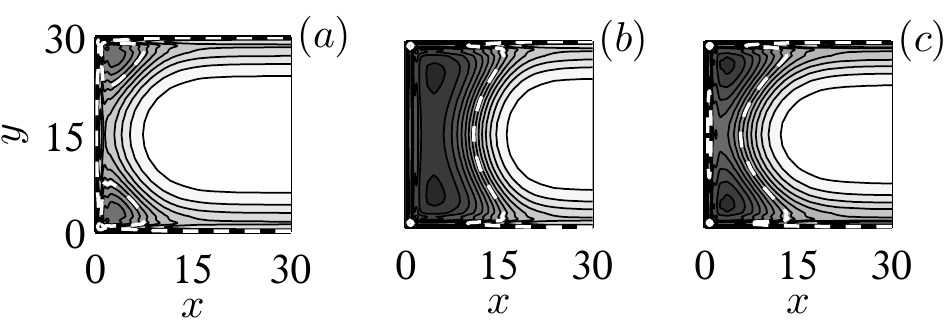}%
    \caption{Density profiles coexisting during capillary prewetting at different values of $T$ in the capillary of figure \ref{Figis1}. (a) and (b) $T=0.95$, $\Delta\mu_{\text{cc}}\left(T\right)=-1.4\cdot10^{-2}$, $\Delta\mu_{\text{cpw}}\left(T\right)=-1.7\cdot10^{-2}$, refrence densities:  $\rho_{\text{cc}}^{\text{vap}}=0.11$, $\rho_{\text{cc}}^{\text{liq}}=0.40$. The isotherms have a single Van der Waals loop and look similar to those in figure \ref{Figis1} (a), (b). However, the coexisting density profile (a) from the vapour branch of $\Omega^{\text{ex}}\left(\Delta\mu\right)$ has a different topology: it shows corner drops. (c) Density profile at the critical capillary prewetting point, at $T_{\text{cpw}}^{\text{cr}}=0.963$, $\Delta\mu_{\text{cpw}}^{\text{cr}}\equiv\Delta\mu_{\text{cpw}}\left(T_{\text{cpw}}^{\text{cr}}\right)=-1.9\cdot10^{-2}$ (see figure \ref{FigLV}); $\Delta\mu_{\text{cc}}\left(T_{\text{cpw}}^{\text{cr}}\right)=-1.4\cdot10^{-2}$, $\rho_{\text{cc}}^{\text{vap}}=0.13$, $\rho_{\text{cc}}^{\text{liq}}=0.37$.\label{Figprof1}}%
\end{figure}

For the purposes of all following discussions we note that a
thermodynamically stable fluid phase can be identified with a concave branch
of the free energy (or excess free energy), as a function of the control
parameter. When a system can be in multiple phases, its free energy has
multiple concave branches. The transition value of the control parameter,
where two (or more) fluid phases coexist, is then defined as the abscissa of
the point where the two (or more) concave branches of free energy intersect.
We consider capillaries immersed in a large reservoir, where the state of the
(bulk) fluid can be controlled. Bulk fluid states are defined by the
temperature, $T$, and chemical potential, $\mu=\mu_{\text{sat}}+\Delta\mu$.
We consider isothermal thermodynamic routes ($T$ is fixed), so the only
control parameter is $\mu$. In what follows we will identify completely a
concave branch of the excess free energy as a function of the chemical
potential, $\Omega^{\text{ex}}\left(\Delta\mu\right)$, with the concept of a
fluid phase.

We will discuss the structure of 2D density profiles,
$\rho^{\text{cpd}}\left(x,y\right)$, in detail in section
\ref{StongerConfinement}, where we consider an example of the density
distribution possessing pronounced excluded volume effects. For now we note
that for the chosen value of $T$, the obtained coexisting profiles are quite
representative of the two fluid surface phases corresponding to the
intersecting stable branches of $\Omega^{\text{ex}}\left(\Delta\mu\right)$,
figure \ref{Figis1}(a). We will refer to them as \emph{vapour} (dashed line,
typical profiles are shown in figures \ref{Figis1}(c) and \ref{Figprof1}(a))
and \emph{capillary-liquid slab} (solid line, typical profiles are shown in
figures \ref{Figis1}(d) and \ref{Figprof1}(b)). As will be made clear in
section \ref{PhDiagr} from the consideration of the full phase diagram, this
first-order phase transition is indicative of a new capillary bulk phase
transition in the same way as, e.g., the wall prewetting (coexistence between
two microscopic liquid films on the surface of a planar wall) is indicative
of first-order wall wetting \cite{Dietrich}. We therefore refer to the
described phase transition as \emph{capillary prewetting}.

For wide capillaries, where the corners are sufficiently isolated, one might
expect a mechanism of wetting related to the filling of corners. Indeed, at
higher temperatures the fluid configurations from the vapour branch of the
excess free energy become denser in the near-corner regions as $\Delta\mu$ is
increased and can even exhibit distinct droplet-like structures with sharp
interfaces. An example is given in figures \ref{Figprof1}(a) and
\ref{Figprof1}(b), where at $T=0.95$ the coexisting density configurations
are those of two drops in the corners and a capillary-liquid slab. However,
the excess free energy and adsorption isotherms look qualitatively similar to
those shown in figure \ref{Figis1} (only two concave branches). The
development of drops in the corners happens continuously from the vapour
phase (at low $\Delta\mu$ the density profiles are of type shown in figure
\ref{Figprof1}(a)), and we cannot associate it with a surface phase
transition here, as there is no additional Van der Waals loop on the wetting
isotherm. However, as we will show in section \ref{StongerConfinement}, there
can be an additional first-order phase transition associated with drop
formation in capillary corners, and it is related to wedge prewetting (see,
e.g. \cite{RejDietNapPRE99}). Finally, we note that a further increase of $T$
leads to criticality of the capillary prewetting transition and the
calculated coexisting density profiles become indistinguishable, see figure
\ref{Figprof1}(c). We will return to this point in much more detail in part
II.


\subsection{Phase diagram}
\label{PhDiagr} 
\begin{figure}
\centering
    \includegraphics{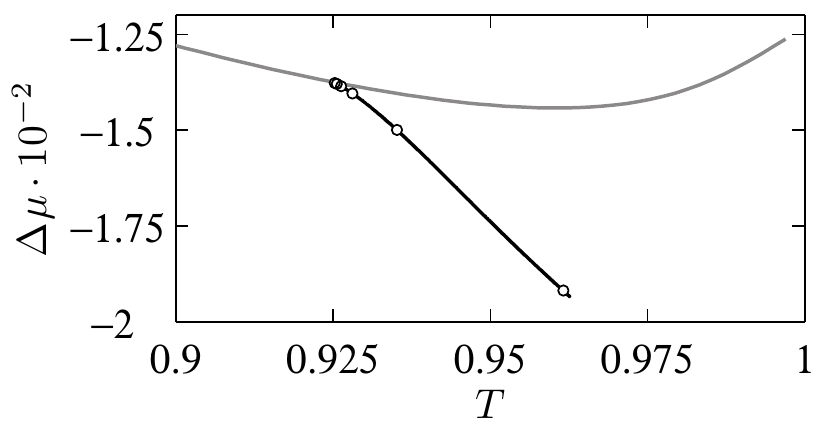}%
    \caption{Full phase diagram of wetting on the capped capillary from figure \ref{Figis1}.
    The CC transition line of capillary bulk (associated slit pore), $\Delta\mu_{\text{cc}}\left(T\right)$,
    is plotted in grey. It ends at the CC critical point,
    $\left(T_{\text{cc}}^{\text{cr}},\Delta\mu_{\text{cc}}^{\text{cr}}\right)\equiv\left(0.996,-1.26\cdot10^{-2}\right)$.
    The transition line of capillary prewetting, $\Delta\mu_{\text{cpw}}\left(T\right)$, is plotted in black and is
    tangential to the CC transition line at $\left(T_{\text{cw}},\Delta\mu_{\text{cw}}\right)\equiv\left(0.925,-1.38\cdot10^{-2}\right)$,
    calculated with accuracy $\Delta_{\text{cw}}=0.0016\cdot10^{-2}$ (see main text). The transition line
    $\Delta\mu_{\text{cpw}}\left(T\right)$ ends at the capillary prewetting critical point:
    $\left(T_{\text{cpw}}^{\text{cr}},\Delta\mu_{\text{cpw}}^{\text{cr}}\right)\equiv\left(0.963,-1.93\cdot10^{-2}\right)$.
    The open circles on $\Delta\mu_{\text{cpw}}\left(T\right)$ denote the capillary prewetting transitions,
    whose coexisting slab profiles are plotted in figure \ref{FigToTw}. \label{FigLV}}%
\end{figure}

To systematise the study of the first-order phase transitions
between near-wall fluid configurations and the continuous transitions to CC,
we obtain the complete phase diagram of the capped capillary, where we plot
the disjoining chemical potential of the transitions versus the transition
temperature. Thus, one wetting isotherm (e.g., figure \ref{Figis1}) possibly
corresponds to one point on the line forming the locus of phase transitions
in the $T$ -- $\Delta\mu$ space (we will refer to them as \emph{transition
lines}). The phase diagram of the capped capillary we considered above is
presented in figure \ref{FigLV} and consists of two transition lines. The CC
transition line of the associated slit pore (capillary bulk),
$\Delta\mu_{\text{cc}}\left(T\right)$ (grey), forms in the case of the capped
capillary the locus of continuous transitions corresponding to the diverging
adsorption (see figure \ref{Figis1}(b)). The transition line of capillary
prewetting, $\Delta\mu_{\text{cpw}}\left(T\right)$ (black), forms the locus
of the crossing branches of $\Omega^{\text{ex}}\left(\Delta\mu\right)$ (see
figure \ref{Figis1}(a)).

Each point on a transition line corresponds to the two coexisting density
profiles $\rho_1$, $\rho_2$ and the chemical potential of the transition
$\mu$, which enter equation \eref{tr} as unknowns. For the
$\Delta\mu_{\text{cc}}\left(T\right)$-line:
$\rho_1\equiv\rho^{\text{slt}}_1\left(y\right)$ and
$\rho_2\equiv\rho^{\text{slt}}_2\left(y\right)$ are the coexisting vapour and
capillary-liquid density profiles inside the slit pore (where CC is a
first-order transition) and
$\Omega^{\text{ex}}\equiv\Omega\left[\rho_y\right]$, as defined by equation
\eref{Om} in the space of 1D fluid configurations, $\rho\left({\bf
r}\right)\equiv\rho^{\text{slt}}\left(y\right)$. Each point on
$\Delta\mu_{\text{cpw}}\left(T\right)$-line, in turn, satisfies the same
equation, where $\rho_1\equiv\rho^{\text{cpd}}_1\left(x,y\right)$ and
$\rho_2\equiv\rho^{\text{cpd}}_2\left(x,y\right)$ are the 2D density
configurations coexisting inside the capped capillary during the first-order
vapour -- slab transition, and $\Omega^{\text{ex}}$ is defined by equation
\eref{OmEx}.

In practice, to find a solution to equation \eref{tr} by, e.g., Newton's
method, one needs to provide an initial guess of the coexisting density
profiles and the transition chemical potential. Equation \eref{tr} is highly
non-linear, non-local, involving integrations over the entire calculation
domain, which makes it quite unstable numerically, and a poor initial guess
results in non-convergence. To find a good initial guess to solve equation
\eref{tr} at a given temperature, one should compute a free energy isotherm
at that temperature (e.g., figure \ref{Figis1}(a)), find by interpolation the
approximate intersection point of its branches, which would provide the
initial guess for $\mu$. The two density profiles, corresponding to the data
points on the excess free energy isotherm and belonging to its vapour and
slab branches, closest to the intersection, can provide the initial guess for
$\rho_1$ and $\rho_2$. In the example from figure \ref{Figis1}, the data
points giving rise to the initial guess for equation \eref{tr} are marked by
open circles on the isotherms. In the case of non-convergence the calculation
of the isotherm should be refined to provide more data points near the
intersection of its branches and thus a better initial guess. After
convergence of the numerical scheme one obtains a single point on a
transition line. The rest of the transition line is best found by arc-length
continuation, using the obtained solution as the starting point.

The dependence of equation \eref{tr} on $T$ parametrizes the set of
first-order phase transitions as a one-dimensional curve in the space spanned
by the two coexisting profiles and the transition chemical potential. Adding
a geometric constraint of curve continuity allows to trace the whole set,
starting from a single point (defined by the two coexisting profiles and the
transition chemical potential) by, e.g., applying the arc-length continuation
to the discretised equation \eref{tr}, where $T$ is treated as the parameter.
The practical value of this approach is quite significant: although any
wetting isotherm gives rise to no more than a single point on each of the
transition lines, we do not need to compute sets of isotherms in order to
obtain those transition lines, but can instead systematically trace each line
with temperature in the same manner as we obtained the isotherms (by
``tracking'' with $\mu$ an easily obtainable density profile in the vapour
phase). The danger of tracking a particular phase transition with $T$ is that
one obtains the transition line of that transition only; the approach is
oblivious to the possible presence of other first-order phase transitions
happening at the same $T$, but possibly at a different $\mu$, which can only
be revealed by a Van der Waals loop on an isotherm. Therefore, after
computing the phase diagram by tracing each transition line individually with
arc-length continuation, we selectively compute several isotherms from the
considered range of temperatures -- to make sure, that there are no
unaccounted phase transition in that range.

Let us consider the phase diagram from figure \ref{FigLV} in more detail. The
CC transition line, $\Delta\mu_{\text{cc}}\left(T\right)$ (grey), forms the
full phase diagram of the slit pore associated with the capped capillary. It
separates the configurations of the capillary filled entirely with vapour
(below the CC line on the phase diagram) from those filled entirely with
capillary-liquid (above the CC line), and is obtained from a fully
microscopic approach, where the only model input is the pairwise fluid-fluid
and fluid-substrate inter-molecular interactions (unlike the empirical
equation \eref{Kelvin}, which provides a macroscopic model of the same
phenomenon).

The capped capillary reduces to its associated slit pore as $x\to\infty$,
thus the fluid in the slit forms the bulk of the capped capillary, and CC is
the \emph{bulk transition}. The $\Delta\mu_{\text{cpw}}\left(T\right)$-line,
which extends into the bulk vapour segment of the phase diagram is one of our
main findings and we examine it in more detail. First, we show that it does
run tangent to the CC line. Second, we show that the point of contact is
associated with a new type of phase transition happening in the capillary
bulk, due to the presence of the capping wall.
\begin{figure}
\centering
    \includegraphics{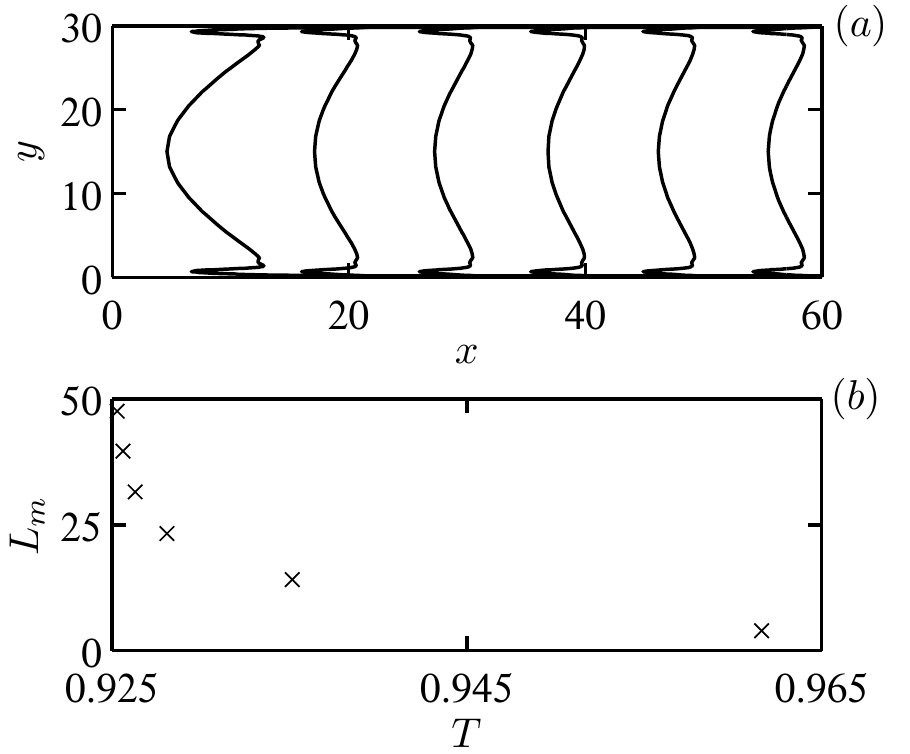}%
    \caption{Liquid menisci of configurations inside the capped capillary with parameters given in
    the caption of figure \ref{Figis1}, coexisting with vapour configurations during capillary prewetting
    transitions at values of $T$ (right to left): 0.9253, 0.9256, 0.9263, 0.9281, 0.9352, 0.9616. The transitions
    at these values of $T$ are marked by open circles on the full phase diagram, figure \ref{FigLV}.
    (b) Menisci heights, $L_{\text{m}}$, for configurations shown in (a) as a function of temperature.
    As $T$ is lowered along the transition line, the length of coexisting capillary-liquid slab increases
    and diverges in the limit  $T\to T_w\approx 0.925$. \label{FigToTw}}%
\end{figure}

Figure \ref{FigToTw}(a) shows the liquid -- vapour interfaces (menisci) of
the fluid configurations with capillary-liquid slabs, coexisting during
capillary prewetting with vapour configurations. On the
$\Delta\mu_{\text{cpw}}\left(T\right)$-line in figure \ref{FigLV}, the
respective transitions are marked by open circles. As the value of $T$ is
lowered, the meniscus of a coexisting slab configuration unbinds into the
capillary bulk. On the phase diagram such fluid configurations fall on the
$\Delta\mu_{\text{cpw}}\left(T\right)$-line, where it approaches the
$\Delta\mu_{\text{cc}}\left(T\right)$-line. We can define the measure of the
``meniscus height'', i.e. the distance from the capping wall to the liquid
meniscus as
\begin{equation}
L_{\text{m}} = \Gamma/\left(H\rho_{\text{cc}}^{\text{liq}}\right).
\end{equation}
As it is clear from figure \ref{FigToTw}(b), where we plot $L_{\text{m}}$ for
the configurations in figure \ref{FigToTw}(a) as a function of $T$, the
menisci heights of coexisting slab configurations diverge along the
$\Delta\mu_{\text{cpw}}\left(T\right)$-line, as it approaches the CC line.
There clearly exists a \emph{limiting} value of temperature,
$T=T_{\text{cw}}$, such that $L_{\text{m}}\to\infty$ as $T\to T_{\text{cw}}$
and the entire capillary is filled with capillary-liquid. Thus, for an
isothermal thermodynamic route to CC at $T=T_{\text{cw}}$, the configuration
where the capillary is filled entirely with vapour coexists with the
configuration possessing a capillary-liquid slab of an infinite length (i.e.,
the capillary is filled entirely with capillary-liquid), which is \emph{only}
possible if the point
$\left(T_{\text{cw}},\Delta\mu_{\text{cpw}}\left(T_{\text{cw}}\right)\right)$
on the phase diagram belongs to \emph{both} transition lines, namely
$\Delta\mu_{\text{cc}}\left(T_{\text{cw}}\right)=\Delta\mu_{\text{cpw}}\left(T_{\text{cw}}\right)$.
Thus, the $\Delta\mu_{\text{cpw}}\left(T_{\text{cw}}\right)$-line approaches
the $\Delta\mu_{\text{cc}}\left(T_{\text{cw}}\right)$ as $T\to
T_{\text{cw}}$, and has a single common point with it at $T=T_{\text{cw}}$.
In other words, $\Delta\mu_{\text{cpw}}\left(T_{\text{cw}}\right)$-line runs
tangent to the CC line.

The phenomenology of wetting on the capped capillary, thus, maps to that of a
fluid in contact with a planar wall and undergoing a first-order wetting
transition: the CC line acts as saturation line, and the
$\Delta\mu_{\text{cpw}}\left(T\right)$-line acts as the wall prewetting line,
see, e.g., \cite{Yatsyshin2012}. Following the analogy with planar wetting,
we have termed the vapour -- slab transition as capillary prewetting, while
the value $T=T_{\text{cw}}$ is termed \emph{capillary wetting temperature}.
Note that our observation for the existence of capillary wetting temperature
relies on the observed divergence of the coexisting slab configurations, as
the CC transition line is approached along the capillary prewetting line. The
presence of the capillary wetting temperature has been confirmed with
numerous calculations performed over a broad range of parameteres and appears
to be the feature of the 2D geometry.

While the capillary wetting temperature, $T_{\text{cw}}$, is associated with
an infinite capillary-liquid slab, for practical purposes the calculations
are always restricted to a finite domain. In the case of planar wetting, one
has access to mean-field critical exponents near the planar wetting
temperature, $T_{\text{w}}$, which can be obtained from thermodynamics
\cite{SaametalJLTP1992,HaugeSchickPhysRevB83}, and used to extrapolate the
calculation data and find the wetting temperature of a particular substrate
with an almost experimental precision, see, e.g.,
\cite{Sart091,ChenEtAlPhysRevLet93}. In our case such analytic results are
obviously unavailable, and we entirely rely on the computations. For every
calculated phase diagram we provide the value of the accuracy,
$\Delta_{\text{cw}}$, for the calculated capillary wetting temperature,
$\bar{T}_{\text{cw}}\approx T_{\text{cw}}$, which we define as
\begin{equation}
\Delta_{\text{cw}}\equiv\left|\Delta\mu_{\text{cpw}}\left(\bar{T}_{\text{cw}}\right)-\Delta\mu_{\text{cc}}\left(\bar{T}_{\text{cw}}\right)\right|.
\end{equation}

The capillary wetting temperature separates the two types of CC. For
isothermal thermodynamic routes to CC at $T<T_{\text{cw}}$ the phase, where
fluid configurations possess a finite capillary-liquid slab, is metastable.
CC then happens discontinuously from the capillary bulk as a first-order
transition, and the 2D capped capillary does not exhibit any principal
difference to a 1D slit pore. On the other hand, at $T>T_{\text{cw}}$ the CC
is a continuous transition: it happens from the capping wall. The
capillary-liquid slab grows continuously into the capillary bulk, as, e.g.,
$\mu\to\mu_{\text{cc}}$ isothermally. If the capillary prewetting line is
crossed (see figure \ref{FigLV}), the CC is preceded by the first-order
transition between fluid configurations with vapour and a capillary-liquid
slab of finite length (e.g., figure \ref{Figis1}). Hence, an isothermal
approach to CC at $T=T_{\text{cw}}$ can be viewed as the limiting case of
capillary prewetting, where the coexisting slab is of infinite length:
$L_{\text{m}}\to\infty$. The discovery of $T_{\text{cw}}$ is of fundamental
importance. Even though the capillary prewetting transition may be influenced
by thermal fluctuations, which in the DF methodology are left unaccounted
for, the presence of prewetting would be revealed by the two distinct types
isothermal CC in the capillary bulk, namely first-order versus continuous,
depending on whether $T$ is below or above $T_{\text{cw}}$. This should be
observable experimentally and is impervious to thermal fluctuations.

The higher-temperature end of the capillary prewetting line at
$T=T_{\text{cpw}}^{\text{cr}}$ is a critical point. Our mean-field DF
approach reveals the signature of criticality: the coexisting configurations
become structurally indistinguishable and the branches of free energy
defining vapour and slab phases align to form a single branch in the limit
$T\to T_{\text{cpw}}^{\text{cr}}$. The critical prewetting density profile
for the capped capillary we have considered in this section is shown in
figure \ref{Figprof1}(c).


\subsection{Fluid structure, connection to wedge wetting}
\label{StongerConfinement}

Here we explore deeper the microscopic surface phases formed near the capping
wall. First, we describe in detail the fluid structure, second, we come back
to the question which was left open in the previous section: can the
mechanism of corner wetting be intensified to possibly form a separate
surface phase? So far we have observed some structural resemblance to wedge
wetting (e.g., the profile with corner drops in figure \ref{Figprof1}(c)),
but the formation of such configurations happened continuously and was not
associated with a dedicated concave branch of excess free energy. Since we
focus now on the microscopic details of fluid structure, a comment on the
role of fluctuations in the system is in order. For capillaries, where the
difference between the capillary wetting temperature, $T_{\text{cw}}$, and
prewetting critical temperature, $T_{\text{cpw}}^{\text{cr}}$, is large, one
can arguably select a value of
$T:T_{\text{cpw}}<T<T_{\text{cpw}}^{\text{cr}}$ to be sufficiently low, so
that the effects of interface fluctuations would be comparable to those on a
planar interface and will not distort significantly the fluid structure.

We proceed to examining a capillary with a weaker substrate potential (lower
planar $T_{\text{w}}$): $\varepsilon_{\text{w}}=0.85$,
$\sigma_{\text{w}}=1.5$, $H_0=2.8$, $H=30$. The fluid is treated in WDA, the
planar wetting temperature being $T_{\text{w}}=0.868$. The capillary wetting
temperature is $T_{\text{cw}}=0.87$, with the accuracy
$\Delta_{\text{cw}}=0.006\cdot10^{-2}$.

We discuss the fluid structure on the example of the density profile with the
capillary-liquid slab. For illustration purposes we would like the
temperature to be low, so that the fluid density has a pronounced structure,
but, at the same time, it should be above $T_{\text{cw}}$ -- for the slab
phase to be thermodynamically stable. We chose the value $T=0.88$, just above
$T_{\text{cw}}$. All analysis below holds for any slab-like configuration
(even metastable) with $L_{\text{m}}$ larger than the near-wall region, where
the density oscillates. For example, in the case of our chosen substrate
potential the meniscus hight should satisfy $L_{\text{m}}>5$.
\begin{figure}
\centering
    \includegraphics{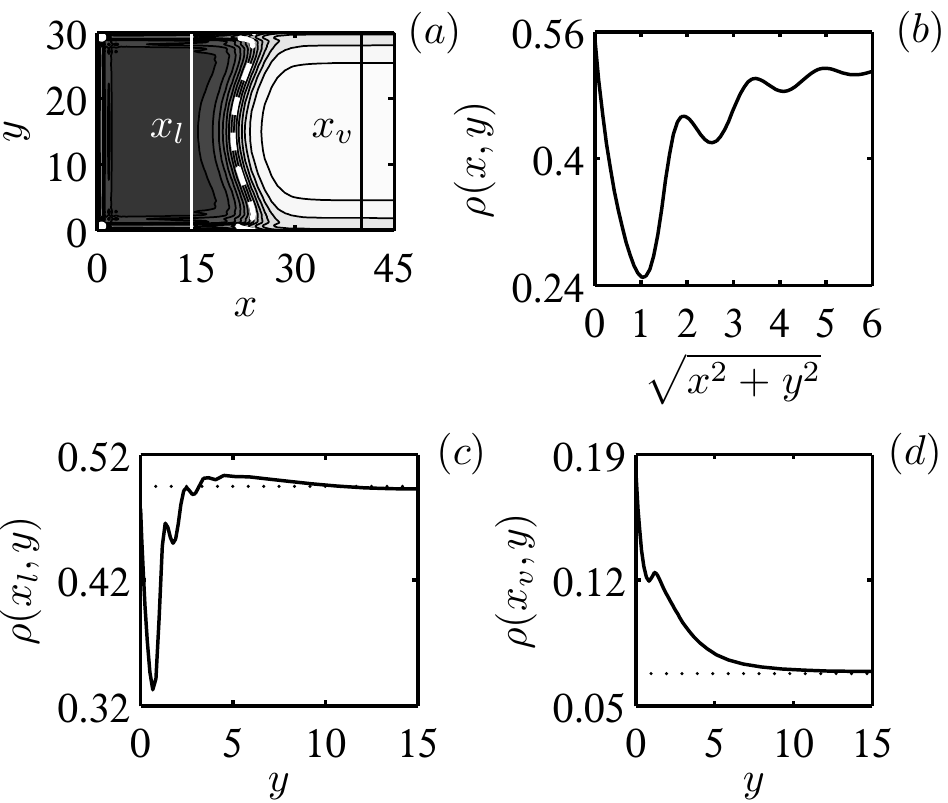}%
    \caption{(a) Density profile of the coexisting capillary-liquid slab configuration at $T=0.88$,
    $\Delta\mu_{\text{cpw}}\left(T\right)=-2.4\cdot10^{-2}$ in the capillary with $H=30$,
    $\varepsilon_{\text{w}}=0.85$, $\sigma_{\text{w}}=1.5$, $H_0=2.8$; the fluid is treated in WDA,
    planar $T_{\text{w}}=0.868$. Vertical lines show the position of slices inside the capillary-liquid,
    at $x_{\text{l}}=14$, and inside the vapour, at $x_{\text{v}}=40$.
    (b) Slice along a bisector. (c), (d) Slices inside capillary-liquid and
    vapour at $x_{\text{l}}$, $x_{\text{v}}$ (due to symmetry, shown for $0\leq y\leq H/2$),
    dotted horizontal lines are at $\rho_{\text{cc}}^{\text{liq}}=0.5$ and $\rho_{\text{cc}}^{\text{vap}}=0.07$. 
    \label{FigSlices}}%
\end{figure}

Generally, the presence of the substrate walls causes oscillations of the
fluid density in the near wall region. The effect is due to the competition
between the attraction of fluid molecules to the wall (which is strongest in
the vicinity of the wall) and the very strong repulsions between the fluid
molecules at short distances. The latter effect is, in turn, due to an
overlap in electron orbitals and the Pauli exclusion principle, and is
included into classical fluid models by prescribing the particles to have
rigid finite-sized cores -- hard spheres of (strictly speaking)
temperature-dependent diameter, which, for reasons given in the introduction,
we set to unity ($\sigma\equiv1$), without loss of generality. The WDA DF
approximation to the fluid free energy, developed from the analysis of the
molecular fluid-fluid correlations \cite{TarazonaWDA0}, treats the repulsions
non-locally and captures this excluded volume effect in the microscopic fluid
structure.

Figure \ref{FigSlices} shows the density profile (figure \ref{FigSlices}(a))
and three representative cross sections: along the corner bisector (figure
\ref{FigSlices}(b)), well inside capillary-liquid (at $x_{\text{l}}$, figure
\ref{FigSlices}(c)) and vapour (at $x_{\text{v}}$, figure
\ref{FigSlices}(d)). The values $x_{\text{l}}$ and $x_{\text{v}}$ are marked
on the density profile by white and black vertical lines. Considering the
slice along the corner bisector in figure \ref{FigSlices}(b), we note
pronounced oscillations, set approximately one hard sphere diameter apart,
with amplitude rapidly (exponentially, \cite{HendersonBook92}) decaying away
from the apex of the corner. The same is true about the slice inside
capillary-liquid in figure \ref{FigSlices}(c). Even along a slice inside
vapour one can see a single oscillation, figure \ref{FigSlices}(d).

The horizontal dotted lines in figures \ref{FigSlices}(c) and
\ref{FigSlices}(d) mark the values of $\rho_{\text{cc}}^{\text{liq}}$ and
$\rho_{\text{cc}}^{\text{vap}}$. The presented slices further uncover the
physics of an isothermal slab growth as $\mu\to\mu_{\text{cc}}$ and
illustrate the reasoning behind our choice of the reference values
$\rho_{\text{cc}}^{\text{vap}}$ and $\rho_{\text{cc}}^{\text{vap}}$ for
scaling the plotted data. As can be seen from the figures, the density slices
inside each phase, $\rho\left(x_{\text{l}},y\right)$ and
$\rho\left(x_{\text{v}},y\right)$, tend to their corresponding reference
values, $\rho_{\text{cc}}^{\text{liq}}$ and $\rho_{\text{cc}}^{\text{vap}}$,
with $y\to\infty$. Moreover, our multiple computations show that, generally,
the density cross sections taken well inside the vapour and slab phases are
\emph{exactly identical} (within the margin of machine rounding error) to the
1D density profiles, $\rho_{y,\text{liq}}\left(y\right)$ and
$\rho_{y,\text{vap}}\left(y\right)$, of fluid configurations coexisting
during CC in the associated slit pore, at the same $T$:
\begin{eqnarray}
\rho\left(x_{\text{l}},y\right)\equiv\rho_{y,\text{liq}}\left(y\right),\nonumber\\ \rho\left(x_{\text{v}},y\right)\equiv\rho_{y,\text{vap}}\left(y\right).
\label{equality}
\end{eqnarray}

The latter result is very important for our proposed analogy in the
phenomenology of wetting on a capped capillary and on a planar wall. In
planar wetting at $T>T_{\text{w}}$, when the nearly saturated fluid
($\mu\lesssim\mu_{\text{sat}}$) is brought in contact with the planar
substrate wall, its density profile exhibits a characteristic plateau of
nearly constant liquid-like density at the wall, see, e.g., reference
\cite{Yatsyshin2012}. The nearly constant value at the plateau is exactly the
density of the liquid at saturation ($\mu=\mu_{\text{sat}}$), at the same
$T$. For wetting on a capped capillary, the role of the bulk is played by the
fluid phase in the associated slit pore. Increasing $\mu$ at
$T>T_{\text{cw}}$ leads to the formation of the liquid-like slab, whose
length diverges in the limit of the continuous CC, as
$\mu\to\mu_{\text{cc}}$. The equivalence expressed by equations
\eref{equality} proves, that the slab is formed from nothing else, but the
capillary-liquid, which coexists with vapour during CC in the slit pore,
associated with the capped capillary. Thus, CC is indeed the bulk transition
for the capped capillary and the phenomenology is identical to that of planar
wetting. Note, however, that wetting on a capped capillary belongs to a
different Ising universality class, than planar wetting, see
\cite{Yatsyshin2013, Parry07}. The different universality class entails,
e.g., that as bulk coexistence is approached in both systems, the divergence
of adsorption in the case of a planar wall and the capped capillary follows
different power laws: for the case of the planar wall
$\Gamma\sim\Delta\mu^{-1/3}$, while in the case of the capped capillary
$\Gamma\sim\left(\mu-\mu_{\text{cc}}\right)^{-1/4}$.

Finally, we note that a slice inside vapour (considered up to the half of
capillary width), $\rho\left(x_{\text{v}},y\right)$, figure
\ref{FigSlices}(c), is also exactly identical to the density profile of a
fluid (vapour) at the same $T$ and $\mu$ in contact with a single planar
wall. In other words, ``vapour'' configurations inside a capped capillary, a
slit pore, and in contact with a planar wall are all \emph{identical}, which
is why the respective phase is everywhere referred to simply as vapour, in
contrast to, e.g., capillary-liquid phase, where the density plateau in a
slit pore is different from that at a planar wall.
\begin{figure}
\centering
    \includegraphics{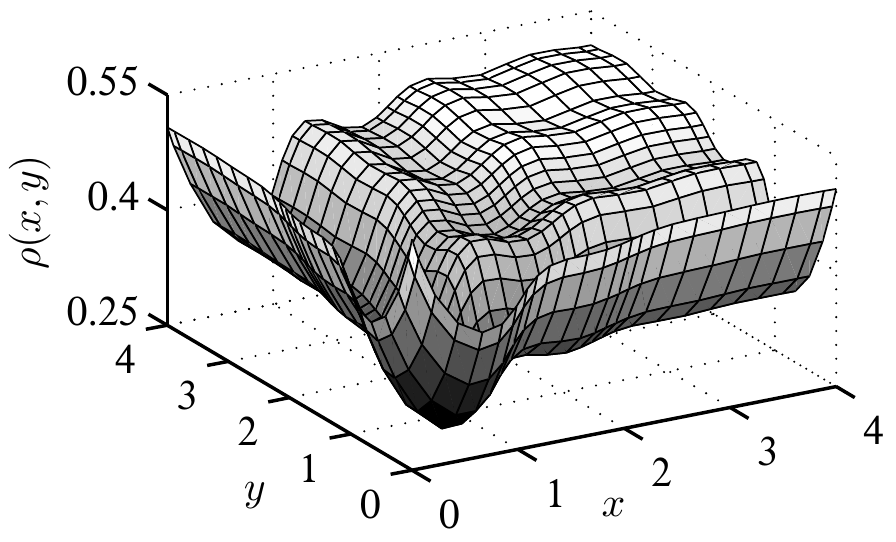}%
    \caption{Surface plot of the density profile from figure \ref{FigSlices}(a) in the near-corner region,
    showing the oscillations of fluid density due to the non-local excluded volume interactions.\label{FigProf3D}}%
\end{figure}

The spatially heterogeneous near-corner fluid structure is best visualized
with a surface plot of $\rho^{\text{cpd}}\left(x,y\right)$ and is presented
in figure \ref{FigProf3D}. Considering it in more detail, we note that the
density oscillations are primarily localized near the apex of the corner and
decay rapidly into the capillary bulk. The absolute maximum of the density is
reached exactly at the corner apex, followed by its absolute minimum
positioned on the corner bisector (see also figure \ref{FigSlices}(b)). In
the presented example the capillary is quite wide, its side walls are far
apart, so the presented part of the $\rho^{\text{cpd}}\left(x,y\right)$
surface looks symmetric about the bisector and resembles the fluid density
distribution inside a symmetric wedge, e.g., \cite{BrykEurphysLett03}.
However, in a narrower capillary that seeming near-corner symmetry would be
broken by the influence of the second side wall, whose total effect would
result from the combination of molecular repulsions due to the increased
geometrical confinement of the fluid and attractions to the wall. By
following standard steps, it is possible to obtain exact sum rules, linking
the values of the 2D density, $\rho^{\text{cpd}}\left(x,y\right)$, at contact
with the capping wall, the side walls, and even the corner apex, with the
bulk thermodynamic variables $T$ and $\mu$, as it was done, e.g. in
references \cite{HendJChemPhys03,HendJChemPhys04} for open wedges immersed in
liquid.
\begin{figure}
\centering
    \includegraphics{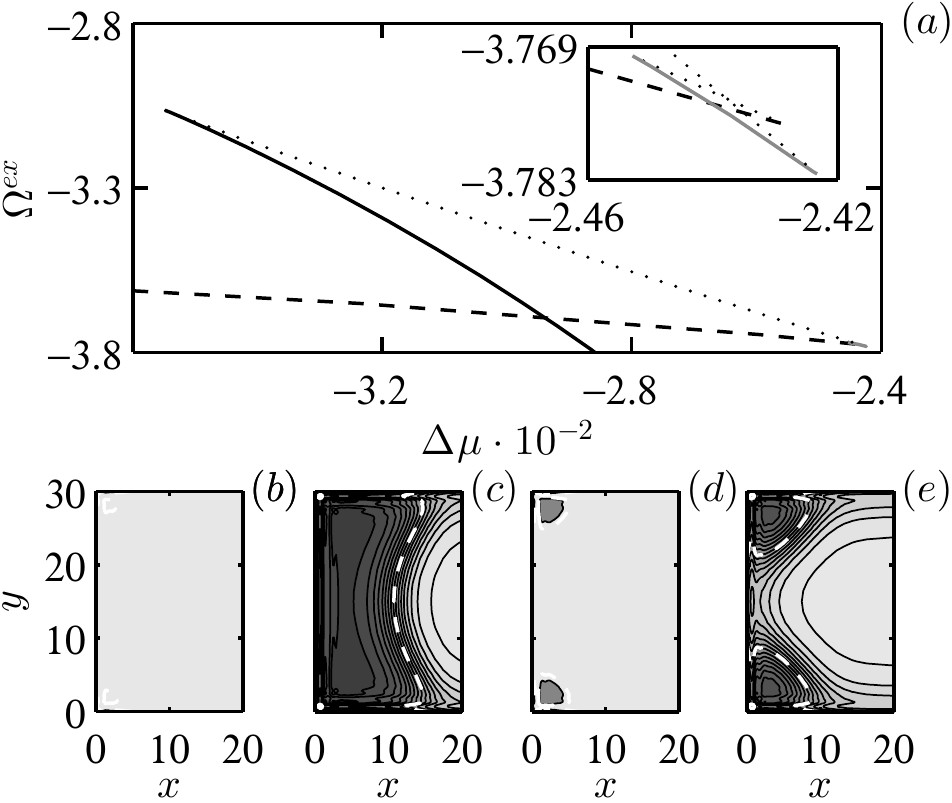}%
    \caption{Isothermal adsorption in capillary from figure \ref{FigSlices} at $T=0.906$;
    $\Delta\mu_{\text{cc}}\left(T\right)=-2.4\cdot10^{-2}$, reference densities:
    $\rho_{\text{cc}}^{\text{vap}}=0.08$, $\rho_{\text{cc}}^{\text{liq}}=0.46$.
    (a) Excess free energy isotherm. It possesses three concave branches
    (connected by non-concave branches, dotted line), which define three fluid phases:
    vapour (dashed), capillary-liquid (solid black) and drops (solid grey, see inset).
    There are two consecutive phase transitions (revealed by intersections of concave branches):
    capillary prewetting, at $\Delta\mu_{\text{cpw}}=-2.94\cdot10^{-2}$, $\Omega^{\text{ex}}=-3.70$,
    and (metastable, see inset) shifted wedge prewetting, at $\Delta\tilde{\mu}_{\text{wpw}}=-2.44\cdot10^{-2}$, $\Omega^{\text{ex}}=-3.77$.
    (b) and (c) Coexisting profiles at capillary prewetting. (d) and (e) Coexisting  profiles at (metastable) shifted wedge prewetting. \label{FigLVDis}}%
\end{figure}

A careful investigation of the fluid phase behaviour reveals that besides the
capillary prewetting, an additional first-order phase transition can occur,
which involves fluid configurations possessing drops in the corners. For the
chosen substrate parameters this transition takes place only between
metastable fluid configurations. A representative excess free energy isotherm
is shown in figure \ref{FigLVDis}(a) and has three concave branches, which
define the fluid phases of vapour (dashed line), capillary-liquid slab (black
solid line) and corner drops (grey solid line, see inset). The isotherm
exhibits two Van der Waals loops, that of capillary prewetting (where dashed
branch crosses the solid black one), and that of the additional (metastable)
phase transition (where the dashed branch crosses the solid grey branch,
magnified on the inset). The density profiles of the two consecutive
transitions are shown in figures \ref{FigLVDis}(b) -- \ref{FigLVDis}(e). Note
that the vapour configuration coexisting with the capillary-liquid slab
(figure \ref{FigLVDis}(b)), possesses a level set at
$\left(\rho_{\text{cc}}^{\text{vap}}+\rho_{\text{cc}}^{\text{liq}}\right)/2$,
noticeable in the capillary corners. The existence of that level set here is
due to the excluded volume effects and the resulting near-wall density
oscillations, unlike, e.g., the configurations with corner drops on plots (d)
and (e), which coexist during the first-order transition, see inset of figure
\ref{FigLVDis}(a).

\begin{figure}
\centering
    \includegraphics{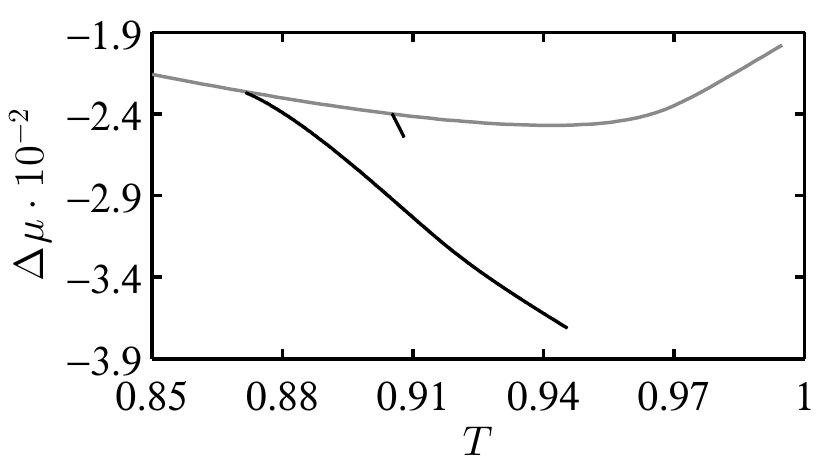}%
    \caption{Phase diagram of the capped capillary from figure \ref{FigSlices}.
    Solid grey curve: CC-line, ending at
    $\left(T_{\text{cc}}^{\text{cr}},\Delta\mu_{\text{cc}}\left(T_{\text{cc}}^{\text{cr}}\right)\right)=\left(0.995,-1.97\cdot10^{-2}\right)$.
    Solid black lines: capillary prewetting, $\Delta\mu_{\text{cpw}}\left(T\right)$,
    which is tangent to CC at
    $\left(T_{\text{cw}},\Delta\mu_{\text{cpw}}\left(T_{\text{cw}}\right)\right)=\left(0.871,-2.23\cdot10^{-2}\right)$, $\Delta_{\text{cw}}=0.0065\cdot10^{-2}$,
    and ends at $\left(T_{\text{cpw}}^{\text{cr}},\Delta\mu_{\text{cpw}}\left(T_{\text{cw}}^{\text{cr}}\right)\right)=\left(0.945,-3.7\cdot10^{-2}\right)$,
    and shifted wedge prewetting, $\Delta\tilde{\mu}_{\text{wpw}}\left(T\right)$, which is bounded by
    $\left(\tilde{T}_0,\Delta\mu_{\text{cpw}}\left(\tilde{T}_0\right)\right)=\left(0.905,-2.4\cdot10^{-2}\right)$ and
    $\left(\tilde{T}_{\text{wpw}}^{\text{cr}},\Delta\mu_{\text{cpw}}\left(\tilde{T}_{\text{wpw}}^{\text{cr}}\right)\right)=\left(0.908,-2.5\cdot10^{-2}\right)$.
    The lower-$T$ end of shifted wedge prewetting belongs to the CC line.
    \label{FigLVD1}}%
\end{figure}

The full phase diagram of the capped capillary is shown in figure
\ref{FigLVD1}. The grey curve is the CC transition line
($\Delta\mu_{\text{cc}}\left(T\right)$), the black curves are the capillary
prewetting transition line ($\Delta\mu_{\text{cpw}}\left(T\right)$, it is
tangential to $\Delta\mu_{\text{cc}}\left(T\right)$) and the locus of
(metastable) transitions of the type shown in the inset of figure
\ref{FigLVDis}(a), taking place between vapour and drop phases. We will
denote it as $\Delta\tilde{\mu}_{\text{wpw}}\left(T\right)$-line. In our
example it is bounded at the lower-$T$ end, at $T_0=0.87$, and at the
higher-$T$ end at $T\equiv\tilde{T}_{\text{wpw}}^{\text{cr}}=0.885$).

First, note that the capillary wetting temperature is lower and the capillary
prewetting spans a broader temperature range, than in the example from the
previous section: for the phase diagram in figure \ref{FigLV} we have
$T_{\text{cw}}^\text{cr}-\bar{T}_{\text{cw}}=0.374$, while for the one in
figure \ref{FigLVD1} we have
$T_{\text{cw}}^\text{cr}-\bar{T}_{\text{cw}}=0.737$. The effect can be
attributed to the relatively weaker attracting substrate. We use the value of
planar wetting temperature, $T_{\text{w}}$, as an effective measure of the
substrate attractive strength, with a lower $T_{\text{w}}$ corresponding to a
weaker substrate. For the capillary in figure \ref{FigLV} we have
$T_{\text{w}}=0.927$, while for the capillary in figure \ref{FigLVD1} the
value is lower: $T_{\text{w}}=0.868$. In planar wetting a weaker attracting
substrate also has a more pronounced prewetting transition line, as shown in,
e.g., reference \cite{Yatsyshin2012}.

Regarding the $\Delta\tilde{\mu}_{\text{wpw}}\left(T\right)$-line we note
that its higher-temperature end at $T=\tilde{T}_{\text{wpw}}^{\text{cr}}$ is
a critical point. Computing various free energy isotherms for
$T\to\left(\tilde{T}_{\text{wpw}}^{\text{cr}}\right)^{-}$ we find, that the
concave branch defining the drop phase tends to align with the branch
defining the vapour phase (see, e.g., dashed and solid grey lines in figure
\ref{FigLVDis}(a)) and forms a single branch in the above limit. The density
profiles of fluid configurations coexisting along the
$\Delta\tilde{\mu}_{\text{wpw}}\left(T\right)$-line become identical in the
same limit. On the other hand, the lower-temperature end of the transition
line at $T=\tilde{T}_0$ does not possess the above signature of criticality:
at $T\lesssim \tilde{T}_0$ excess free energy isotherms still possess three
concave branches, with the branch defining the drop phase being entirely
metastable, without an intersection with the vapour branch.


In the previous section we have noted the existence of fluid configurations
possessing corner drops (see coexisting profiles in figures \ref{Figprof1}(a)
and \ref{Figprof1}(b)). Although the nature of the phenomenon was unclear,
the relation to wetting on a wedge was prompted by the fluid structure. On
the other hand, we have established that a phase diagram of the capped
capillary can possess a separate transition line,
$\Delta\tilde{\mu}_{\text{wpw}}\left(T\right)$, forming the locus of
first-order transitions between fluid configurations with corner drops. We
also found, that the $\Delta\tilde{\mu}_{\text{wpw}}\left(T\right)$-line ends
at a critical temperature, $\tilde{T}_{\text{wpw}}^{\text{cr}}$, above which
the formation of corner drops happens continuously with increasing
$\Delta\mu$ at a fixed $T$. So, by considering again the continuous formation
of the corner drops in the capillary from the previous section (figures
\ref{Figprof1}(a) and \ref{Figprof1}(b)), we see that somehow the effect of
criticality at $\tilde{T}_{\text{wpw}}^{\text{cr}}$ is still present, even
though the transition line (which should end at
$T=\tilde{T}_{\text{wpw}}^{\text{cr}}$) is not found on the full phase
diagram (figure \ref{FigLV}). Such effect is only possible, if the transition
to configurations with corner drops (given in the case of figure
\ref{FigLVD1} by the $\Delta\tilde{\mu}_{\text{wpw}}\left(T\right)$-line) is
not related to the capped capillary, but exists \emph{independently}. Based
on these arguments, we now can relate the transition line
$\Delta\tilde{\mu}_{\text{wpw}}\left(T\right)$ to the wedge prewetting (whose
transition line we would denote $\Delta\mu_{\text{wpw}}\left(T\right)$),
which is studied in detail in, e.g., reference \cite{RejDietNapPRE99}.

In Part II of this study we will show, that the entire wedge prewetting line,
$\Delta\mu_{\text{wpw}}\left(T\right)$, and its critical temperature,
$T_{\text{wpw}}^{\text{cr}}$, are shifted in the capped capillary due to
confinement, analogously to how bulk coexistence is shifted in a slit pore
due to the Kelvin effect, see equation \eref{Kelvin}. In the case of the
capped capillary discussed in section \ref{secLV} (phase diagram in figure
\ref{FigLV}), the coexisting fluid configurations from figures
\ref{Figprof1}(a) and \ref{Figprof1}(b) are supercritical with respect to the
shifted wedge prewetting. We expect, that either the
$\Delta\mu_{\text{wpw}}\left(T\right)$-line ends above the
$\Delta\mu_{\text{cc}}\left(T\right)$-line, or the shift of
$\Delta\mu_{\text{wpw}}\left(T\right)$-line, due to the second side wall of
the capillary, does not allow for $\Delta\tilde{\mu}\left(T\right)$-line to
form.

To better understand the interplay between wedge and capillary wetting
through the DF-type simulation, one can compute the full wedge prewetting
line, and then study, how that line is influenced, by adding a third wall
(second side wall of the capped capillary) at various distances $H$, and
computing the phase diagrams of the resulting capped capillaries of varying
heights. Such an exploration, however, goes beyond the scope of the present
study dedicated to the study of capped capillaries. 

\begin{figure}
\centering
    \includegraphics{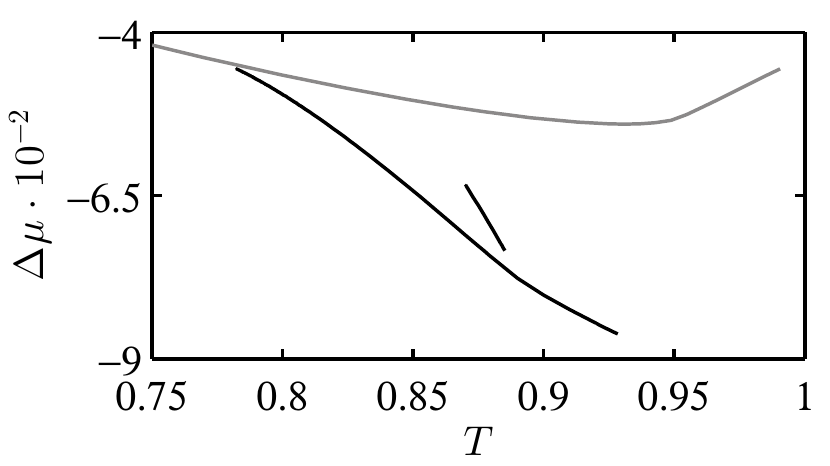}%
    \caption{Phase diagram of the capped capillary with $\varepsilon_{\text{w}}=0.7$, $\sigma_{\text{w}}=2$, $H_0 = 5$, $H=30$;
    fluid treated in LDA, planar $T_{\text{w}}=0.755$. Solid grey curve:
    CC-line, ending at $\left(T_{\text{cc}}^{\text{cr}},\Delta\mu_{\text{cc}}\left(T_{\text{cc}}^{\text{cr}}\right)\right)=\left(0.99,-4.56\cdot10^{-2}\right)$.
    Solid black lines: capillary prewetting, $\Delta\mu_{\text{cpw}}\left(T\right)$, which is tangent to CC at
    $\left(T_{\text{cw}},\Delta\mu_{\text{cpw}}\left(T_{\text{cw}}\right)\right)=\left(0.78,-4.6\cdot10^{-2}\right)$, $\Delta_{\text{cw}}=0.06\cdot10^{-2}$,
    and ends at $\left(T_{\text{cpw}}^{\text{cr}},\Delta\mu_{\text{cpw}}\left(T_{\text{cw}}^{\text{cr}}\right)\right)=\left(0.928,-8.62\cdot10^{-2}\right)$,
    and shifted wedge prewetting, $\Delta\tilde{\mu}_{\text{wpw}}\left(T\right)$, which is bound by
    $\left(\tilde{T}_0,\Delta\mu_{\text{cpw}}\left(\tilde{T}_0\right)\right)=\left(0.87,-6.33\cdot10^{-2}\right)$ and
    $\left(\tilde{T}_{\text{wpw}}^{\text{cr}},\Delta\mu_{\text{cpw}}\left(\tilde{T}_{\text{wpw}}^{\text{cr}}\right)\right)=\left(0.885,-7.34\cdot10^{-2}\right)$.
    Note that lower-$T$ end of shifted wedge prewetting line does not lie on the CC transition line, unlike, e.g., figure \ref{FigLVD1}.
\label{FigLVD2}}%
\end{figure}

Finally, we note that the lower-temperature end of the shifted wedge
prewetting line (at $\tilde{T}_0$) does not have to coincide with the CC
line, as can be see from, e.g., figure \ref{FigLVD2}, which shows the full
phase diagram of the capped capillary with parameters
$\varepsilon_{\text{w}}=0.7$, $\sigma_{\text{w}}=2$, $H_0 = 5$, $H=30$; fluid
is treated in LDA, planar wetting temperature: $T_{\text{w}}=0.755$. The
system is similar to the one we have studied in reference
\cite{Yatsyshin2013}.


\section{Summary}

Our study of wetting phenomena in capped capillaries is based on statistical
mechanics, in particular non-local interactions in the fluid were taken into
account explicitly through the use of DF formulation for the fluid free
energy. We employed two DF approximations for the fluid free energy: one
treating repulsions non-locally (WDA) and one treating them locally (LDA).
The fact, that both approximations treat attractions in a non-local fashion
and lead to essentially the same wetting behaviour of the fluid (above the
bulk triple point, in the range of liquid- and vapour-like densities), shows
that the presented physics of wetting is controlled by the long-ranged
attractive forces acting in the fluid, see, e.g., figures \ref{FigLVD1} and
\ref{FigLVD2}.

We have implemented an arc-length continuation technique to address two very
important problems of mean-field analyses of physical systems, namely the
computation of an isotherm (adsorption and free energy), and the computation
of a phase diagram. The methodology is readily applicable to other fluid
settings while the most important step in its practical implementation is
obtaining a starting point. Typically, one can consider some limiting value
of the continuation parameter, where the equation is numerically
well-behaved, i.e., one does not have to come up with an initial guess for
the Newton iterations to converge to a solution. An example is provided in
the calculation of free-energy or an adsorption isotherm, where the
continuation is initiated from a low value of the parameter, $\mu$, hence the
system is in the vapour phase, and almost unaffected by the substrate. The
Newton algorithm converges fast and a simple initial guess
$\rho^{\text{cpd}}\left(x,y\right)\equiv\rho_{\text{cc}}^{\text{vap}}$
suffices. On the other hand, the computation of a phase line involves the
analysis of an approximate isotherm to get a starting point, but the value of
the continuation parameter ($T$) need not correspond to any special limiting
case of the problem.

For the study of the fluid phase behaviour we have adopted a method, where we
identify the fluid phase with a concave branch of the excess free energy in
the space of thermodynamic fields. A first-order phase transition then is
given by the intersection of concave branches. The order parameter for the
transitions we presented is the adsorption: a first-order transition is
associated with a finite jump in its value, while a continuous transition --
with its divergence. Our results can be summarised as follows:

\begin{itemize}
\item We have shown the existence of the capillary wetting temperature,
    $T_{\text{cw}}$, which is the intrinsic property of the 2D pore, in
    the same sense as the wetting temperature, $T_{\text{cw}}$, is the
    property of the 1D planar substrate. It controls the order of the
    phase transition in capillary bulk: at $T<T_{\text{cw}}$ CC is of
    first order (as it is in a slit pore), while at $T>T_{\text{cw}}$ it
    is a continuous phase transition associated with the diverging slab
    of capillary-liquid formed at the capping wall. Obtaining the
    capillary wetting temperature requires one to solve a set of 2D
    Euler-Lagrange equations in the space of density profiles
    $\rho\left(x,y\right)$ for various values of temperature and chemical
    potential.
\item We have shown the existence of the first-order capillary prewetting
    transition, where the vapour coexists with a capillary-liquid slab of
    a finite length formed at the capping wall. The CC at
    $T=T_{\text{cw}}$ can be viewed as the limiting case of capillary
    prewetting. The typical phase diagram of the capped capillary, thus,
    consists of the phase diagram of the slit pore, which acts as
    capillary bulk, and the transition line of capillary prewetting,
    which is the consequence of 2D confinement of the fluid.
\item The remnant of wetting on wedge-like substrates is manifested in
    capped capillaries by a (metastable) fluid phase with drops in the
    capillary corners. In Part II we shall demonstrate, that in the
    capped capillary the wedge prewetting is shifted, due to confinement
    by the additional side wall.
\item The fluid has been found to possess a highly oscillatory structure
    in the corners. The oscillations are due to the excluded volume
    effects and decay rapidly away from the apices.
\end{itemize}

We expect our findings to stimulate further theoretical research of
microscopic pores as well as experimental investigations. In Part II we
explore deeply the complicated interplay between different mechanisms of 2D
wetting. We consider an example of the system, which exhibits two types of
three-phase coexistence: first, the capped capillary allows for stable
vapour, drop and capillary-liquid slab phases, second, the capillary bulk
allows for stable phases of vapour, planar prewetting film and
capillary-liquid. We also identify potentially fruitful areas for future
investigation, which are prompted by our results presented in both parts.

\section*{Acknowledgements}
We are grateful to Prof. S. Dietrich for a fruitful discussion on wedge
wetting and to the European Research Council via Advanced Grant No. 247031
for support of this research.

\section*{References}

\end{document}